\documentclass{SciPost}

\usepackage[english]{babel}

\usepackage[utf8]{inputenc} 
\usepackage{lmodern} 
\usepackage[T1]{fontenc}

\usepackage{amssymb}
\usepackage{mathtools} 

\usepackage{bbm} 

\usepackage{verbatim} 
\usepackage{xcolor}

\usepackage{graphicx}

\usepackage{tikz}
\usetikzlibrary{calc} 
\usetikzlibrary{decorations.pathreplacing}
\usetikzlibrary{decorations.pathmorphing} 
\usetikzlibrary{matrix,arrows} 
\usetikzlibrary{fadings} 

\DeclarePairedDelimiter{\bra}{\langle}{\rvert}
\DeclarePairedDelimiter{\ket}{\lvert}{\rangle}
\DeclarePairedDelimiterX{\ketbra}[2]{\lvert}{\rvert}{#1\rangle \langle#2}
\DeclarePairedDelimiterX{\braket}[2]{\langle}{\rangle}{#1\vert#2}

\DeclarePairedDelimiterX{\cbraket}[2]{\langle\!\langle}{\rangle}{#1\vert#2}
\DeclarePairedDelimiterX{\bracket}[2]{\langle}{\rangle\!\rangle}{#1\vert#2}
\DeclarePairedDelimiterX{\cbracket}[2]{\langle\!\langle}{\rangle\!\rangle}{#1\vert#2}

\usepackage{bm} 
\newcommand{\vect}[1]{\bm{{#1}}}

\newcommand{\I}{\mathrm{i}}
\newcommand{\E}{\mathrm{e}}

\newcommand{\bd}{\begin{displaymath}}
	\newcommand{\ed}{\end{displaymath}}
\newcommand{\be}{\begin{equation}}
	\newcommand{\ee}{\end{equation}}
\newcommand{\bea}{\begin{eqnarray}}
	\newcommand{\eea}{\end{eqnarray}}

\newcommand{\R}{\mathbb{R}}

\DeclareMathOperator*{\ordprod}{\prod\limits^{\vbox to -.5ex{\kern-0.5ex\hbox{$\leftharpoonup$}\vss}}}
\DeclareMathOperator*{\ordprodopp}{\prod\limits^{\vbox to -.5ex{\kern-0.5ex\hbox{$\rightharpoonup$}\vss}}}

\makeatletter
\DeclareRobustCommand\widecheck[1]{{\mathpalette\@widecheck{#1}}}
\def\@widecheck#1#2{%
	\setbox\z@\hbox{\m@th$#1#2$}%
	\setbox\tw@\hbox{\m@th$#1%
		\widehat{%
			\vrule\@width\z@\@height\ht\z@
			\vrule\@height\z@\@width\wd\z@}$}%
	\dp\tw@-\ht\z@
	\@tempdima\ht\z@ \advance\@tempdima2\ht\tw@ \divide\@tempdima\thr@@
	\setbox\tw@\hbox{%
		\raise\@tempdima\hbox{\scalebox{1}[-1]{\lower\@tempdima\box
				\tw@}}}%
	{\ooalign{\box\tw@ \cr \box\z@}}}
\makeatother

\hypersetup{
    colorlinks,
    linkcolor={red!50!black},
    citecolor={blue!50!black},
    urlcolor={blue!80!black}
}

\usepackage[bitstream-charter]{mathdesign}
\DeclareSymbolFont{usualmathcal}{OMS}{cmsy}{m}{n}
\DeclareSymbolFontAlphabet{\mathcal}{usualmathcal}

\fancypagestyle{SPstyle}{
\fancyhf{}
\lhead{\colorbox{scipostblue}{\bf \color{white} ~SciPost Physics }}
\rhead{{\bf \color{scipostdeepblue} ~Submission }}

\fancyfoot[C]{\textbf{\thepage}}
}

\begin{document}

\pagestyle{SPstyle}

\begin{center}
	{\Large \textbf{\color{scipostdeepblue}{
	The deformed Inozemtsev spin chain
	}}}
\end{center}

\begin{center}\textbf{
Rob Klabbers\textsuperscript{1\,$\star$},
and
Jules Lamers\textsuperscript{2$\,\curvearrowleft\,$3\,$\dagger$}
}\end{center}

\begin{center}
\textbf{1} Humboldt-Universität zu Berlin \\
Zum Großen Windkanal 2, 12489 
Berlin, Germany
\\[.5\baselineskip]
\textbf{2} Université Paris--Saclay, CNRS, CEA \\ Institut de Physique Théorique \\ 91191 Gif-sur-Yvette, France
\\[.5\baselineskip]
\textbf{3} Deutsches Elektronen-Synchrotron DESY \\ Notkestraße 85, 22607 Hamburg, Germany 
\\[\baselineskip]
$\star$ \href{mailto:rob.klabbers@physik.hu-berlin.de}{\small rob.klabbers@physik.hu-berlin.de}\,,\quad
$\dagger$ \href{mailto:jules.lamers@desy.de}{\small jules.lamers@desy.de}
\end{center}

\section*{\color{scipostdeepblue}{Abstract}}
\boldmath
\textbf{The Inozemtsev chain is an exactly solvable interpolation between the short-range Hei\-senberg and long-range Haldane--Shastry (HS) chains. In order to unlock its potential to study spin interactions with tunable interaction range using the powerful tools of integrability, the model's	mathematical properties require better understanding. As a major step in this direction, we present a new generalisation of the Inozemtsev chain with spin symmetry reduced to $\textit{U}(1)$, interpolating  between a Heisenberg \textsc{xxz} chain and the \textsc{xxz}-type HS chain, and integrable throughout. Underlying it is a new quantum many-body system that extends the elliptic Ruijsenaars system by including spins, contains the trigonometric spin-Ruijsenaars--Macdonald system as a special case, and yields our spin chain by `freezing'. Our models have potential applications from condensed-matter to high-energy theory, and provide a crucial step towards a general theory for long-range integrability. 
}
\unboldmath

\vspace{\baselineskip}

\tableofcontents

\section{Introduction}

Recent years brought tremendous progress for trapped-ion and cold-atom experiments, and low-dimensional systems with tunable spin-spin interactions can now be engineered~\cite{jurcevic2014quasiparticle, zhang2017observation,RevModPhys.93.025001,zeiher2017coherent}.  Wheareas such systems inherently have \emph{long-range} spin interactions, theoretical studies often assume drastically simplified nearest-neighbour interactions. Long-range spin interactions also find applications in quantum information and computing \cite{PhysRevX.11.031016,linke2017experimental,10.1063/1.5088164} and pose fundamental questions about e.g.\ causality  \cite{defenu2023long,PhysRevLett.109.025303,PhysRevLett.111.207202,gong2016kaleidoscope}. In $1+1$ dimensions, \emph{(quantum) integrable} models are exactly solvable thanks to underlying symmetries. Such models may thus offer exciting opportunities to study the effects of long-range interactions using exact analytical methods. Yet such models are rare, and the theory behind them is incomplete.

\paragraph{Main results.} We introduce two new integrable long-range models with spins:

\begin{tabular}{cl}
	\!\!\!\!\!\!\!
	\tikz[baseline={([yshift=-11pt*.3]current bounding box.center)},xscale=.2,yscale=.3]{
		\draw[very thin,gray!50] (0,0) ellipse (4 and 1);
		\draw[dotted] (1*60:4 and 1) -- (4*60:4 and 1);
		\foreach \th in {0*60,2*60,3*60,5*60} { 
			\node at ($(\th:4 and 1)-(0,.1)$) {\tikz[scale=.2]{\fill [black] (0,0) circle (6pt);}};
			\draw [thick,->] ($(\th:4 and 1)-(0,.5)$) -- ($(\th:4 and 1)+(0,.65)$);
		};
		\foreach \th in {1*60,4*60} {
			\node at ($(\th:4 and 1)-(0,.1)$) {\tikz[scale=.2]{\fill [black] (0,0) circle (6pt);}};
			\draw [thick,<-] ($(\th:4 and 1)-(0,.65)$) -- ($(\th:4 and 1)+(0,.5)$);
		};
	} & 
	a (quantum) spin chain; \\
	\!\!\!\!\!\!\!
	\tikz[baseline={([yshift=-11pt*.3]current bounding box.center)},xscale=.2,yscale=.3]{
		\draw[very thin,gray!40] (0,0) ellipse (4 and 1);
		\draw[dotted] (1*60+12:4 and 1) -- ($(4*60+8:4 and 1)$);
		\foreach \th in {0*60-10,2*60-12,3*60+20,5*60+15} { 
			\node at ($(\th:4 and 1)-(0,.1)$) {\tikz[scale=.2]{\fill [black] (0,0) circle (6pt);}};
			\draw [thick,->] ($(\th:4 and 1)-(0,.5)$) -- ($(\th:4 and 1)+(0,.65)$);
		};
		\foreach \th in {1*60+12,4*60+8} {
			\node at ($(\th:4 and 1)-(0,.1)$) {\tikz[scale=.2]{\fill [black] (0,0) circle (6pt);}};
			\draw [thick,<-] ($(\th:4 and 1)-(0,.65)$) -- ($(\th:4 and 1)+(0,.5)$);
		};
		\def\gsinit{55};
		\draw[gray!\gsinit] (0*60-10+36:4.04 and 1.03) arc [start angle = 0*60-10+36, end angle=0*60-10+50, x radius = 4.04, y radius = 1.03];
		\draw[gray!\gsinit] (1*60+12-5:4.04 and 1.03) arc [start angle = 1*60+12-5, end angle=1*60+12-15, x radius = 4.04, y radius = 1.03];
		\draw[gray!\gsinit] (2*60-12+5:4.04 and 1.03) arc [start angle = 2*60-12+5, end angle=2*60-12+15, x radius = 4.04, y radius = 1.03];
		\draw[gray!\gsinit] (3*60+20+12:4.04 and 1.03) arc [start angle = 3*60+20+12, end angle=3*60+20+28, x radius = 4.04, y radius = 1.03];
		\draw[gray!\gsinit] (4*60+8+6:4.04 and 1.03) arc [start angle = 4*60+8+6, end angle=4*60+8+30, x radius = 4.04, y radius = 1.03];
		\draw[gray!\gsinit] (5*60+15-7:4.04 and 1.03) arc [start angle = 5*60+15-7, end angle=5*60+15-17, x radius = 4.04, y radius = 1.03];
		\def\gsmid{75};
		\draw[gray!\gsmid] (0*60-10+40:4.04 and 1.03) arc [start angle = 0*60-10+40, end angle=0*60-10+52, x radius = 4.04, y radius = 1.03];
		\draw[gray!\gsmid] (1*60+12-8:4.04 and 1.03) arc [start angle = 1*60+12-8, end angle=1*60+12-17, x radius = 4.04, y radius = 1.03];
		\draw[gray!\gsmid] (2*60-12+8:4.04 and 1.03) arc [start angle = 2*60-12+8, end angle=2*60-12+17, x radius = 4.04, y radius = 1.03];
		\draw[gray!\gsmid] (3*60+20+18:4.04 and 1.03) arc [start angle = 3*60+20+18, end angle=3*60+20+30, x radius = 4.04, y radius = 1.03];
		\draw[gray!\gsmid] (4*60+8+13:4.04 and 1.03) arc [start angle = 4*60+8+13, end angle=4*60+8+32, x radius = 4.04, y radius = 1.03];
		\draw[gray!\gsmid] (5*60+15-10:4.04 and 1.03) arc [start angle = 5*60+15-10, end angle=5*60+15-19, x radius = 4.04, y radius = 1.03];
		\def\gsend{110};
		\draw[gray!\gsend,->,>=stealth] (0*60-10+45:4.04 and 1.03) arc [start angle = 0*60-10+45, end angle=0*60-10+55, x radius = 4.04, y radius = 1.03];
		\draw[gray!\gsend,->,>=stealth] (1*60+12-12:4.04 and 1.03) arc [start angle = 1*60+12-12, end angle=1*60+12-20, x radius = 4.04, y radius = 1.03];
		\draw[gray!\gsend ,->,>=stealth] (2*60-12+12:4.04 and 1.03) arc [start angle = 2*60-12+12, end angle=2*60-12+20, x radius = 4.04, y radius = 1.03];
		\draw[gray!\gsend ,->,>=stealth] (3*60+20+23:4.04 and 1.03) arc [start angle = 3*60+20+23, end angle=3*60+20+33, x radius = 4.04, y radius = 1.03];
		\draw[gray!\gsend ,->,>=stealth] (4*60+8+25:4.04 and 1.03) arc [start angle = 4*60+8+25, end angle=4*60+8+35, x radius = 4.04, y radius = 1.03];
		\draw[gray!\gsend ,->,>=stealth] (5*60+15-14:4.04 and 1.03) arc [start angle = 5*60+15-14, end angle=5*60+15-22, x radius = 4.04, y radius = 1.03];
	} & \!\!\!\!\!
	\begin{tabular}{l} 
		a quantum many-body system (QMBS), \\ 
		of particles with spins \emph{moving} on a circle.
	\end{tabular} \\
\end{tabular}
\medskip

\noindent 
As we shall see, the two models are closely related. Besides having potential applications in both condensed-matter and high-energy theory, our models shed light on 
the three-decade old open problem to understand the integrability of the Inozemtsev chain.

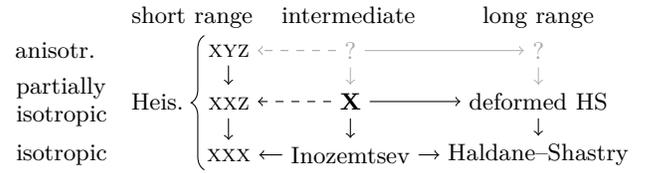
\begin{figure}
	\centering
	\begin{tikzpicture}[nodes={anchor=center}]
		\matrix (m) [matrix of nodes,row sep=-.05em,column sep=.5em]{
			& \!\!\!short range & \vphantom{hg}intermediate\, & long range \\
			\ \ \ \,anisotropic\ \ & \!\!\!\!\hphantom{Heisenberg}\quad \textsc{xyz} & \textcolor{gray!70}{?} & \textcolor{gray!70}{?} \\
			\!\!\begin{tabular}{l} partially \\[-.ex] isotropic \end{tabular} & \!\!\!\!Heisenberg\quad \textsc{xxz} & \textbf{X} & deformed HS \\
			isotropic\ \ & \!\!\!\!\hphantom{Heisenberg}\quad \textsc{xxx} & Inozemtsev & Haldane--Shastry \\
		};
		\draw[decorate,decoration={brace,amplitude=5pt,mirror}] ([xshift=.6cm,yshift=-.1cm]m-2-2.north) -- ([xshift=.6cm,yshift=.1cm]m-4-2.south);
		\draw[dashed,gray!70,->] (m-2-3) -- (m-2-2); \draw[gray!60,->] (m-2-3) -- (m-2-4);
		\draw[dashed,->] (m-3-3) -- (m-3-2); \draw[->] (m-3-3) -- (m-3-4);
		\draw[->] (m-4-3) -- (m-4-2); \draw[->] (m-4-3) -- (m-4-4);
		\draw[->] ([xshift=.95cm]m-2-2.south) -- ([xshift=.95cm]m-3-2.north); 
		\foreach \j in {3,4} \draw[gray!60,->] (m-2-\j) -- (m-3-\j);
		\draw[->] ([xshift=.95cm]m-3-2.south) -- ([xshift=.95cm]m-4-2.north); 
		\foreach \j in {3,4} \draw[->] (m-3-\j) -- (m-4-\j);
	\end{tikzpicture}
	\caption{Landscape of integrable long-range spin chains, including the Heisenberg and Haldane--Shastry chains and their partially isotropic extensions. We find the spot marked `\textbf{X}'.}
	\label{fg:spin_limits}
\end{figure}

\paragraph{The spin chain.}
Until recently, the study of integrable long-range spin chains focused on \emph{isotropic} (i.e.\ $\mspace{-1mu}\mathit{SU}(2)$-symmetric) models. Using a bar to denote isotropic case, these spin chains have hamiltonian of the form
\begin{equation} \label{eq:iso}
	\bar{H} = \frac{1}{2} \sum_{i<j}^N \bar{V}(i-j) \, \bigl(1-\vec{\sigma}_i \cdot \vec{\sigma}_{\!j} \bigr) = \sum_{i<j}^N \bar{V}(i-j) \, \bigl(1-P_{ij}\bigr) \,,
\end{equation}
where we consider a chain of $N$ spins, $\bar{V}(x)$ is a pair potential setting the interaction range, $\vec{\sigma} = (\sigma^x,\sigma^y,\sigma^z)$ are the Pauli spin matrices, and $P_{ij} = (1 + \vec{\sigma}_i \cdot \vec{\sigma}_{\!j})/2$ is the spin permutation operator. The Haldane--Shastry (HS) chain \cite{haldane1988exact,shastry1988exact} 
is given by \eqref{eq:iso} with pair potential
\begin{equation} \label{eq:pot_HS}
	\bar{V}^\textsc{hs}(x) = \frac{1}{r^2} \, , \quad r = \tfrac{N}{\pi} \mspace{2mu} \sin\bigl|\tfrac{\pi}{N}\, x\bigr| \, , 
\end{equation}
which is the critical case for long-range order (cf.\ \cite{PhysRev.187.732,defenu2023long,PhysRevLett.111.207202}). It can be engineered with trapped ions \cite{grass2014trapped} and is a lattice toy model for the fractional quantum Hall effect~\cite{Hal_91a,Hal_94} and Wess--Zumino--Witten CFT~\cite{HH+_92,bernard1994spinons,bouwknegt1994spinon,bouwknegt1996n}. This model is connected (Fig.~\ref{fg:spin_limits}) to the nearest-neighbour Heisenberg \textsc{xxx} chain through the Inozemtsev chain~\cite{Inozemtsev:1989yq}, 
whose hamiltonian $\bar{H}^\text{Ino}$ is given by \eqref{eq:iso}
with 
\begin{equation} \label{eq:pot_Ino}
	\bar{V}^\text{Ino}(x) = \wp(x) + \text{cst} 
\end{equation}
the Weierstraß elliptic function. This pair potential generalises \eqref{eq:pot_HS} by including a second, imaginary period that sets the interaction range. Widely believed to be integrable~\cite{Hal_94,klabbers2022coordinate}, $\bar{H}^\text{Ino}$ offers the tantalising possibility to study a spin system analytically as one tunes the interaction range. First, however, the toolkit of integrability needs to be developed further: there is a conjecture for a hierarchy of conserved charges of $\bar{H}^\text{Ino}$~\cite{inozemtsev1996invariants, dittrich2008commutativity},%
\footnote{\ \textit{Note added during the revisions.} Chalykh \cite{Chalykh24u} has at long last settled this problem by deriving extensively many commuting charges for the elliptic spin-Calogero--Sutherland system and, by `freezing', the Inozemtsev chain. It is not yet clear how the charges of  \cite{inozemtsev1996invariants, dittrich2008commutativity} fit in. The proof of \cite{Chalykh24u} exploits the algebraic framework of elliptic Dunkl operators. Whereas their simpler (trigonometric) counterpart moreover allows one to construct a Yangian underpinning the Haldane--Shastry chain \cite{bernard1993yang}, unfortunately this is not true in the elliptic setting. Explicitly featuring \textit{R}-matrices, the $q$-deformed Inozemtsev chain is an important theoretical asset to help identify a quantum-group-like structure underlying the Inozemtsev chain.}
but no underlying algebraic structure is known. This is an important open problem in the theory of integrability \cite{Hal_94}. To unveil such structures we shall break the spin symmetry of $\bar{H}^\text{Ino}$ in a controlled way.

The HS chain has a \emph{partially isotropic} (i.e.\ $\mspace{-2mu}\mathit{U}(1)$-sym\-metric) extension retaining its key properties, the \emph{deformed} HS 
chain \cite{Ugl_95u,Lam_18,lamers2022spin}. Our first new long-range model likewise deforms $\bar{H}^\text{Ino}$, generalising the Inozemtsev and deformed HS chains as in Fig.~\ref{fg:spin_limits} while remaining integrable.
The partially isotropic generalisation of $1-P_{ij}$ from \eqref{eq:iso} comes in two `chiralities', with \emph{deformed permutations} transporting either spin to the other, for a \emph{deformed exchange}, followed by transport back. Like in \eqref{eq:iso}, a \emph{potential} sets the interaction range; it is a `point splitting' of \eqref{eq:pot_Ino} as anticipated in \cite{klabbers2022coordinate}.

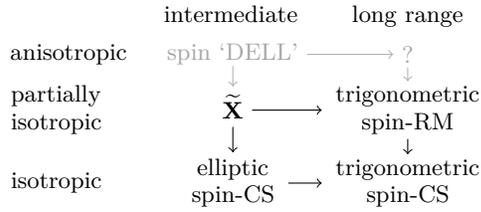
\begin{figure}[t]
	\centering
	\begin{tikzpicture}[nodes={anchor=center}]
		\matrix (m) [matrix of nodes,row sep=.05em,column sep=1em]{
			& \vphantom{hg}intermediate\, & long range \\
			\ \ \: anisotropic \ & \textcolor{gray!70}{spin `DELL'} & \textcolor{gray!70}{?} \\
			\!\!\begin{tabular}{l} partially \\ isotropic \end{tabular} & \vphantom{$\mathbf{X}$}$\smash{\widetilde{\mathbf{X}}}$ & \begin{tabular}{c} trigonometric \\ spin-RM \end{tabular} \\
			isotropic\ \ & elliptic spin-CS & \begin{tabular}{c} trigonometric\\ spin-CS \end{tabular} \\
		};
		\draw[gray!60,->] (m-2-2) -- (m-2-3);
		\draw[->,shorten >=-.15cm] (m-3-2) -- (m-3-3);
		\draw[->,shorten >=-.15cm] (m-4-2) -- (m-4-3);
		\draw[gray!60,->,shorten <=-.1cm,shorten >=.09cm] (m-2-2) -- (m-3-2);
		\draw[gray!60,->,shorten <=-.05cm,shorten >=-.1cm] (m-2-3) -- (m-3-3);
		\draw[->,shorten >=-.075cm] (m-3-2) -- (m-4-2);
		\draw[->,shorten <=-.1cm,shorten >=-.1cm] (m-3-3) -- (m-4-3);
	\end{tikzpicture}
	\caption{Landscape of integrable QMBS with spins, including Calogero--Sutherland (CS) and Ruijsenaars--Macdonald (RM). Without lattice spacing as an infrared cutoff, the short-range limit is absent. We find the spot marked `$\widetilde{\mathbf{X}}$'.}
	\label{fg:qmbs_limits}
\end{figure}

\paragraph{The QMBS.}
Unlike for nearest-neighbour models, integrability of long-range spin chains hinges on connections to QMBSs of Calogero--Sutherland (CS) and Ruijsenaars type. This is best understood for HS (see also \cite{lamers2022fermionic}): 
\begin{enumerate}
	\item[i.] its exact wavefunctions 
	come from a \emph{spinless}
	trigonometric CS system~\cite{Hal_91a,bernard1993yang},
	\item[ii.] its conserved charges stem from a trigonometric CS system \emph{with spins} 
	by `freezing'~\cite{Pol_93,bernard1993yang,TH_95},
\end{enumerate}
and the enhanced (Yangian) spin symmetry of $\bar{H}^\textsc{hs}$ arises from (ii) too~\cite{drinfel1986degenerate,bernard1993yang}.  These connections persist at the partially isotropic level, where trigonometric CS is generalised to the `relativistic' trigonometric Ruijsenaars--Macdonald (RM) system~\cite{bernard1993yang,Ugl_95u,lamers2022spin} (Fig.~\ref{fg:qmbs_limits}). For $\bar{H}^\text{Ino}$ only (i) was properly understood, via the \emph{elliptic} CS system \cite{Inozemtsev_1995,klabbers2022coordinate}. Here, we add (ii): our spin chain arises by freezing an \emph{elliptic dynamical spin-Ruijsenaars system}. This QMBS is our second new long-range model (Fig.~\ref{fg:qmbs_limits}). 
Despite its supporting role here, it is clearly of independent theoretical interest. 
We shall prove the commutativity of its hamiltonians elsewhere.

\paragraph{Outline.}
In Section \ref{sec:spin_chain}, we introduce our new long-range spin chain, discuss how it satisfies the defining properties introduced above, and compute 
two new limits: an intermediate refinement of the Inozemtsev chain, and the short-range limit.
We furthermore point out some interesting new features. In Section \ref{sec:qmbs}, we construct a novel QMBS and discuss its properties. We moreover outline how `freezing' this QMBS yields our spin chain, thereby connecting the commutativity of their respective charges and hence their integrability. We conclude in Section \ref{sec:conclusion}. The appendices contain all relevant information about the elliptic functions (Appendix \ref{app:ell}) and \textit{R}-matrix (Appendices \ref{app:def_perm}--\ref{app:def_nn_exchange}) that we will need. 
\bigskip

\noindent While we focus on spin~$1/2$, all our results extend to higher rank: multi-component versions with several particle `species' (`colours').%
\footnote{\ Simply replace \eqref{eq:Rdyn} by the dynamical $\mathfrak{gl}_r$ $R$-matrix \cite{felder1997elliptic}.}

\section{The spin chain}
\label{sec:spin_chain}

\subsection{Hamiltonians}

Consider $N$ spin-$1/2$ sites equispaced on a circle.
The deformed Inozemtsev chain has `chiral' hamiltonians 
\begin{equation} \label{eq:hamiltonian}
	H^\textsc{l} = \sum_{i<j}^N V\mspace{-1mu}(i-j) \, S_{[i,j]}^\textsc{l} \, , \quad 
	H^\textsc{r} = \sum_{i<j}^N V\mspace{-1mu}(i-j) \, S_{[i,j]}^\textsc{r} \, .
\end{equation}
Let $\rho(x) = \theta'(x)/\theta(x)$, where $\theta(x)$ is the odd Jacobi theta function with quasiperiods 
$\I\pi/\kappa$ and $N$, which we view as a periodic version of a hyperbolic sine:
\begin{equation} \label{eq:theta}
	\theta(x) = \frac{\sinh(\kappa \, x)}{\kappa} \prod_{n=1}^{\infty}\! \frac{\sinh [\kappa \, (N \, n + x)] \, \sinh[\kappa \, (N \, n - x)]}{\sinh^2 (N \kappa \, n)} = \frac{\sinh(\kappa \, x)}{\kappa} + O(p^2) \, ,
\end{equation}
with nome $p = \E^{-N\kappa}$, see Appendix \ref{app:ell} for more.

The potential is
\begin{equation} \label{eq:pot}
	\!\! V\mspace{-1mu}(x) = \frac{\rho(x-\eta) - \rho(x+\eta)}{\theta(2\eta)} 
	\sim \frac{1}{\mathrm{sn}(x+\eta)\,\mathrm{sn}(x-\eta)} \, ,
\end{equation}
with anisotropy parameter $\eta$. Here $\mathrm{sn}$ is the Jacobi elliptic sine function, and in `$\sim$' we omit some constants, see \eqref{eq:pot_app} for the precise relation.

The long-range spin interactions $S_{[i,j]}^\textsc{l}$ and $S_{[i,j]}^\textsc{r}$ are deformations of the isotropic long-range spin exchange interaction $E_{ij} = (1-P_{ij})/2 = (1-\vec{\sigma}_{\!i} \cdot \vec{\sigma}_{\!j})/4$ in \eqref{eq:iso}. The latter admits two `chiral' decompositions into nearest-neighbour steps:
\begin{equation} \label{eq:istropic_chiral_red_decomp}
	\begin{aligned}
		E_{ij} & = P_{j-1\mspace{-1mu},\mspace{1mu}j} \cdots P_{i+1,i+2} \, E_{i,i+1} \, P_{i+1,i+2} \cdots P_{j-1\mspace{-1mu},\mspace{1mu}j} \\
		& = P_{i,i+1} \cdots P_{j-2\mspace{-1mu},\mspace{1mu}j-1} \, E_{j-1\mspace{-1mu},\mspace{1mu}j} \, P_{j-2\mspace{-1mu},\mspace{1mu}j-1} \cdots P_{i,i+1}\, .
	\end{aligned}
\end{equation}
The structure on the right-hand side persists to the partially isotropic level, with suitable replacements for both the spin permutation $P$ and the nearest-neighbour spin interaction $E$. These are both built from Felder's dynamical \textit{R}-matrix \cite{felder1994elliptic}
\begin{equation} \label{eq:Rdyn}
	\check{R}(x,a) = 
	\begin{pmatrix}
		\,1 & \color{gray!80}{0} & \color{gray!80}{0} & \color{gray!80}{0}\, \\
		\,\color{gray!80}{0} & f(\eta,x,\eta\,a) & f(x,\eta,\eta\,a) & \color{gray!80}{0}\, \\
		\,\color{gray!80}{0} & f(x,\eta,-\eta\,a) & f(\eta,x,-\eta\,a) & \color{gray!80}{0}\, \\
		\,\color{gray!80}{0} & \color{gray!80}{0} & \color{gray!80}{0} & 1\,
	\end{pmatrix} ,
	\quad
	f(x,y,z) = \frac{\theta(x)\,\theta(y+z)}{\theta(x+y)\,\theta(z)} \, ,
\end{equation}
depending on a `dynamical' parameter $a$. It satisfies the dynamical Yang--Baxter equation, see Appendix~\ref{app:def_perm}. The deformed spin permutation is
\begin{equation} \label{eq:deformed_permutation}
	P_{i,i+1}(x) = \check{R}_{i,i+1}\bigl(x,a-(\sigma^z_1+\dots+\sigma^z_{i-1})\bigr) = 
	\tikz[baseline={([yshift=-.5*11pt*0.3]current bounding box.center)},xscale=0.4,yscale=0.2,font=\footnotesize]{
		\draw[->] (0,0) -- (0,3);
		\foreach \x in {-1,...,1} \draw (.75+.2*\x,1.5) node{$\cdot\mathstrut$};	
		\draw[->] (1.5,0) -- (1.5,3);
		\draw[rounded corners=2pt,->] (3.5,0) node[below, yshift=.05cm]{$\,x''$} -- (3.5,1) -- (2.5,2) -- (2.5,3) node[above, yshift=-.03cm]{$\vphantom{x}\smash{x''}\,$};
		\draw[rounded corners=2pt,->] (2.5,0) node[below, yshift=.05cm]{$x'$} -- (2.5,1) -- (3.5,2) -- (3.5,3) node[above, yshift=-.03cm]{$\,\smash{x'}\vphantom{x}$};
		\draw[->] (4.5,0) -- (4.5,3);
		\foreach \x in {-1,...,1} \draw (5.25+.2*\x,1.5) node{$\cdot\mathstrut$};	
		\draw[->] (6,0) -- (6,3);
		\node at (-.5,1.5) {$a$};
	} \ , \quad x = x' - x'' \, ,
\end{equation}
where the $i$th and $i+1$st spins cross, carrying along their `inhomogeneity' parameters $x',x''$. The dynamical parameter $a$ is shifted by the spin-$z$ to the left of the \textit{R}-matrix. On the usual spin basis labelled by $s_j$ equal to $\uparrow \equiv +1$ or $\downarrow \equiv -1$ for each $1\leqslant j\leqslant N$ this means
\begin{equation} 
	\begin{aligned}
	\! P_{i,i+1}(x) \, \ket{s_1,\dots,s_N} = {} & \ket{s_1,\dots,s_{i-1}} \nonumber \\ 
	& \otimes \check{R}\bigl(x,a-\textstyle\sum_{k=1}^{i-1} s_k\bigr)\, \ket{s_i,s_{i+1}} \nonumber \\
	& \otimes \ket{s_{i+1},\dots,s_N} \, , 
	\end{aligned}
\end{equation}
so, for example, $P_{23}(x) = \ketbra{\uparrow}{\uparrow} \, \otimes \, \check{R}(x,a-1) \; + \; \ketbra{\downarrow}{\downarrow} \, \otimes\, \check{R}(x,a+1)$.
The properties of these deformed spin permutations are collected in Appendix \ref{app:def_perm}.

Finally, the deformed nearest-neighbour spin exchange is defined from \eqref{eq:deformed_permutation} as
\begin{equation} \label{eq:deformed_exchange}
	E_{i,i+1}(x) = \frac{1}{\theta(\eta) \, V\mspace{-1mu}(x)} \, P_{i,i+1}(-x) \, P_{i,i+1}'(x) = 
	\tikz[baseline={([yshift=-.5*11pt*0.3]current bounding box.center)},xscale=0.4,yscale=0.2,font=\footnotesize]{
		\draw[->] (0,0) -- (0,3);
		\foreach \x in {-1,...,1} \node at (.75+.2*\x,1.5) {$\cdot\mathstrut$};	
		\draw[->] (1.5,0) -- (1.5,3);
		\draw[->] (2.5,0) node[below, yshift=.05cm]{$x'$} -- (2.5,3) node[above, yshift=-.03cm]{$\smash{x'}\vphantom{x}$};
		\draw[->] (3.5,0) node[below, yshift=.05cm]{$\ x''$} -- (3.5,3) node[above, yshift=-.03cm]{$\ \vphantom{x}\smash{x''}$};
		\draw[style={decorate, decoration={zigzag,amplitude=.5mm,segment length=1mm}}] (2.5,1.5) -- (3.5,1.5);
		\draw[->] (4.5,0) -- (4.5,3);
		\foreach \x in {-1,...,1} \node at (5.25+.2*\x,1.5) {$\cdot\mathstrut$};	
		\draw[->] (6,0) -- (6,3);
		\node at (-.5,1.5) {$a$};
	} \ , \quad x = x' - x'' \, .
\end{equation}
This definition, in which we factor out the potential~\eqref{eq:pot}, is chosen such that both 
\eqref{eq:pot} and \eqref{eq:deformed_exchange} have the appropriate limits, as we will see in Section~\ref{sec:properties_limits}. 
The explicit $4\times 4$ matrix determining \eqref{eq:deformed_exchange} is given in Appendix \ref{app:def_nn_exchange}.
Unlike the potential, it depends on $a$.
While the dependence on $x$ is new compared to the Inozemtsev and deformed HS chains, this feature is shared by the elliptic long-range spin chain of Matushko and Zotov \cite{MZ_23b,klabbers2024landscapes}, as well as in all degenerations thereof.

Together, the deformed permutation \eqref{eq:deformed_permutation} and deformed exchange \eqref{eq:deformed_exchange} define the chiral long-range spin interactions $S_{[i,j]}^\textsc{l}$ and $S_{[i,j]}^\textsc{r}$ diagrammatically as
\begin{equation} \label{eq:S^LR_diagr}
	S_{[i,j]}^\textsc{l} = \tikz[baseline={([yshift=-.5*11pt*0.2-1pt]current bounding box.center)},xscale=0.4,yscale=0.2,font=\footnotesize]{
		\draw[->] (10.5,0) node[below]{$N$} -- (10.5,10) node[above]{$N$};
		\draw[->] (9,0) -- (9,10);
		\foreach \x in {-1,...,1} \draw (9.25+.2*\x,-1) node{$\cdot\mathstrut$};
		\foreach \x in {-1,...,1} \draw (9.25+.2*\x,11) node{$\cdot\mathstrut$};
		\draw[rounded corners=2pt,->] (8,0) node[below]{$j$} -- (8,1.5) -- (5,4.5) -- (5,5.5) -- (8,8.5) -- (8,10) node[above]{$\smash{j}$};
		\draw[rounded corners=2pt,->] (7,0) -- (7,1.5) -- (8,2.5) -- (8,7.5) -- (7,8.5) -- (7,10); 
		\draw[rounded corners=2pt,->] (6,0) -- (6,2.5) -- (7,3.5) -- (7,6.5) -- (6,7.5) -- (6,10);
		\foreach \x in {-1,...,1} \draw (6+.2*\x,-1) node{$\cdot\mathstrut$};
		\foreach \x in {-1,...,1} \draw (6+.2*\x,11) node{$\cdot\mathstrut$};
		\draw[rounded corners=2pt,->] (5,0) -- (5,3.5) -- (6,4.5) -- (6,5.5) -- (5,6.5) -- (5,10); 
		\draw[->] (4,0) node[below]{$i$} -- (4,10) node[above]{$i$};
		\draw[->] (3,0) -- (3,10);
		\foreach \x in {-1,...,1} \draw (2.75+.2*\x,-1) node{$\cdot\mathstrut$};
		\foreach \x in {-1,...,1} \draw (2.75+.2*\x,11) node{$\cdot\mathstrut$};
		\draw[->] (1.5,0) node[below]{$1$} -- (1.5,10)  node[above]{$1$};
		\draw[style={decorate, decoration={zigzag,amplitude=.5mm,segment length=1mm}}] (4,5) -- (5,5);
		\foreach \x in {-1,...,1} \draw (2.25+.2*\x,5) node{$\cdot\mathstrut$};
		\foreach \x in {-1,...,1} \draw (9.75+.2*\x,5) node{$\cdot\mathstrut$};	
		\node at (1,5) {$a$};
	} \! , \quad
	S_{[i,j]}^\textsc{r} = \tikz[baseline={([yshift=-.5*11pt*0.2-1pt]current bounding box.center)},xscale=-0.4,yscale=0.2,font=\footnotesize]{
		\draw[->] (10.5,0) node[below]{$1$} -- (10.5,10)  node[above]{$1$};
		\draw[->] (9,0) -- (9,10);
		\foreach \x in {-1,...,1} \draw (9.25+.2*\x,-1) node{$\cdot\mathstrut$};
		\foreach \x in {-1,...,1} \draw (9.25+.2*\x,11) node{$\cdot\mathstrut$};
		\draw[rounded corners=2pt,->] (8,0) node[below]{$i$} -- (8,1.5) -- (5,4.5) -- (5,5.5) -- (8,8.5) -- (8,10) node[above]{$i$};
		\draw[rounded corners=2pt,->] (7,0) -- (7,1.5) -- (8,2.5) -- (8,7.5) -- (7,8.5) -- (7,10);
		\draw[rounded corners=2pt,->] (6,0) -- (6,2.5) -- (7,3.5) -- (7,6.5) -- (6,7.5) -- (6,10);
		\foreach \x in {-1,...,1} \draw (6+.2*\x,-1) node{$\cdot\mathstrut$};
		\foreach \x in {-1,...,1} \draw (6+.2*\x,11) node{$\cdot\mathstrut$};
		\draw[rounded corners=2pt,->] (5,0) -- (5,3.5) -- (6,4.5) -- (6,5.5) -- (5,6.5) -- (5,10);
		\draw[->] (4,0) node[below]{$j$} -- (4,10) node[above]{$\smash{j}$};
		\draw[->] (3,0) -- (3,10);
		\foreach \x in {-1,...,1} \draw (2.75+.2*\x,-1) node{$\cdot\mathstrut$};
		\foreach \x in {-1,...,1} \draw (2.75+.2*\x,11) node{$\cdot\mathstrut$};
		\draw[->] (1.5,0) node[below]{$N$} -- (1.5,10) node[above]{$N$};
		\draw[style={decorate, decoration={zigzag,amplitude=.5mm,segment length=1mm}}] (4,5) -- (5,5);
		\foreach \x in {-1,...,1} \draw (2.25+.2*\x,5) node{$\cdot\mathstrut$};
		\foreach \x in {-1,...,1} \draw (9.75+.2*\x,5) node{$\cdot\mathstrut$};	
		\node at (11,5) {$a$};
	} \! .
\end{equation}
Here each site $1\leqslant k\leqslant N$ has a fixed inhomogeneity parameter $x_k^\star = k$, where the `$\mspace{2mu}{}^\star\mspace{2mu}$' serves to distinguish these fixed parameters from their unrestricted counterparts $x_k$ that will appear in \textsection\ref{sec:qmbs}.
Thus, the deformed long-range spin interactions~\eqref{eq:S^LR_diagr} read
\begin{align} 
	\label{eq:S^L}
	S_{[i,j]}^\textsc{l} = {} & P_{j-1\mspace{-1mu},\mspace{1mu}j}(1) \cdots P_{i+1\mspace{-1mu},\mspace{1mu}i+2}(j-i-1) \; E_{i,i+1}(i-j) \; P_{i+1\mspace{-1mu},\mspace{1mu}i+2}(i-j+1) \cdots P_{j-1\mspace{-1mu},\mspace{1mu}j}(-1) \, , \\[.5\baselineskip]
	\label{eq:S^R}
	S_{[i,j]}^\textsc{r} = {} & P_{i,i+1}(1) \cdots P_{j-2,\mspace{1mu}j-1}(j-i-1) \; E_{j-1\mspace{-1mu},\mspace{1mu}j}(i-j) \; P_{j-2,\mspace{1mu}j-1}(i-j+1) \cdots P_{i,i+1}({-}1) \, ,
\end{align}
in clear analogy to the first and second line, respectively, of the decompositions in \eqref{eq:istropic_chiral_red_decomp}.  

\paragraph{Examples.} At $N=3$ 
the chiral long-range spin interactions read
\begin{equation}
	\begin{gathered} 
	S_{[1,2]}^\textsc{l} = E_{12}(-1) \, , \quad
	S_{[2,3]}^\textsc{l} = E_{23}(-1) \, , \quad
	S_{[1,3]}^\textsc{l} = P_{23}(1) \; E_{12}(-2) \; P_{23}(-1) \, , \\
	S_{[1,2]}^\textsc{r} = E_{12}(-1) \, , \quad
	S_{[2,3]}^\textsc{r} = E_{23}(-1) \, , \quad
	S_{[1,3]}^\textsc{r} = P_{12}(1) \; E_{23}(-2) \; P_{12}(-1) \, .
	\end{gathered}
\end{equation}
For higher $N$ the first few terms look exactly the same, with dependence on $N$ residing in the real (quasi)period of the entries of $E_{i,i+1}(x)$ and $P_{i,i+1}(x)$. At $N=4$ we further need
\begin{subequations}
	\begin{gather}
	\begin{gathered} 
	S_{[3,4]}^\textsc{l} = E_{34}(-1) \, , \quad
	S_{[2,4]}^\textsc{l} = P_{34}(1) \; E_{23}(-2) \; P_{34}(-1) \, , \\ 
	S_{[1,4]}^\textsc{l} = P_{34}(1) \, P_{23}(2) \; E_{12}(-3) \; P_{23}(-2) \, P_{34}(-1) \, , 
	\end{gathered}
\intertext{for the left-chiral interactions, and for the right-chiral ones}
	\begin{gathered} 
	S_{[3,4]}^\textsc{r} = E_{34}(-1) \, , \quad
	S_{[2,4]}^\textsc{r} = P_{23}(1) \, E_{34}(-2) \, P_{23}(-1) \, , \\ 
	S_{[1,4]}^\textsc{r} = P_{12}(1) \, P_{23}(2) \, E_{34}(-3) \, P_{23}(-2) \, P_{12}(-1) \, .
	\end{gathered}
	\end{gather} 
\end{subequations}
In general one obtains $S_{[i,j]}^\textsc{l}$ from $S_{[1,j-i+1]}^\textsc{l}$ by shifting all subscripts $k,k+1$ to $k+i-1,k+i$. The same holds for $S_{[i,j]}^\textsc{r}$.
Note that the $S_{[i,j]}^\textsc{l}$ have the same structure as in \cite{Lam_18} and the $S_{[i,j]}^\textsc{r}$ look like in \cite{lamers2022spin}, the difference being the choice of \textit{R}-matrix.

\subsection{Properties and limits} \label{sec:properties_limits}

While the hamiltonians \eqref{eq:hamiltonian} are more complex than in the isotropic case~\eqref{eq:iso}, their ingredients have clear physical meanings: a potential~\eqref{eq:pot}, a deformed permutation~\eqref{eq:deformed_permutation}, and a deformed spin exchange~\eqref{eq:deformed_exchange}. There are four parameters: the length~$N \geqslant 2$, $\kappa>0$ tuning the interaction range, the anisotropy $\eta$, and the dynamical parameter~$a$.%
\footnote{\ 
These parameters have some constraints, since the potential~\eqref{eq:pot} has poles at $2\eta = N k + \I\pi\, l/\kappa$ for $k,l \in \mathbb{Z}$, and the entries of \eqref{eq:Rdyn} have poles at $\eta\,a = N \, k + \I\pi \, l/\kappa$.}
While the hamiltonians are not hermitian for the standard scalar product, numerics shows that the spectrum is real if $\eta$ is imaginary (i.e.\ the regime $|\Delta|>1$ for the Heisenberg \textsc{xxz} spin chain) and $a$ real.

\paragraph{Defining properties.} The chain \eqref{eq:hamiltonian} contains the Inozemtsev and 
deformed HS chains as in Fig.~\ref{fg:spin_limits}, and is integrable. Let us explain.

When $\eta\to0$ we retrieve the isotropic Inozemtsev hamiltonian $H^\text{Ino}$ given by \eqref{eq:iso} and \eqref{eq:pot_Ino}.
Indeed, \eqref{eq:pot} becomes $-\rho'(x) = \bar{V}^\text{Ino}(x)$ from \eqref{eq:pot_Ino}, and both \eqref{eq:S^L}--\eqref{eq:S^R} yield $1-P_{ij}$ up to a conjugation that is removed by $a\to-\I\mspace{2mu}\infty$, since then $P_{i,i+1}(x) \to P_{i,i+1}$ and (see \textsection\ref{app:def_nn_exchange}) $E_{i,i+1}(x) \to 1-P_{i,i+1}$.

At $\kappa = 0$ we find the deformed HS chain, again up to a conjugation that disappears if $a$ is removed. To see this use $\theta(x) \to N \, \sin(\pi\,x/N)/\pi$ as $\kappa \to 0$. The potential \eqref{eq:pot} thus has long-range limit $V^\text{tri}(x) = (\frac{\pi}{N})^2/ \sin[\frac{\pi}{N}(x+\eta)] \sin[\frac{\pi}{N}(x-\eta)]$. 
When $\kappa\to0$ and moreover $\eta\,a\to-\I\mspace{2mu}\infty$ for fixed $\eta$, the exchange \eqref{eq:deformed_exchange} becomes independent of $x$, namely
\begin{equation} \label{eq:TL}
	E^\text{tri} = 
	\begin{pmatrix}
		\,0 & \color{gray!80}{0} & \color{gray!80}{0} & \color{gray!80}{0}\, \\[-2pt]
		\,\color{gray!80}{0} & \hphantom{+}q^{-1}\! & -q\hphantom{+} & \color{gray!80}{0}\, \\
		\,\color{gray!80}{0} & -q^{-1}\! & q & \color{gray!80}{0}\, \\
		\,\color{gray!80}{0} & \color{gray!80}{0} & \color{gray!80}{0} & 0\,
	\end{pmatrix} 
	, \quad q = \E^{\pi\I\mspace{1mu}\eta/N} \, ,
\end{equation}
acting at sites $i,i+1$. The deformed
permutation~\eqref{eq:deformed_permutation}
reduces to the operator
\begin{equation} \label{eq:R_tri}
	\check{R}^\text{tri}(x) = 1 - \frac{\sin(\tfrac{\pi}{N} x)}{\sin[\tfrac{\pi}{N}(x+\eta)]} \, E^\text{tri}
\end{equation}
at sites $i,i+1$. 
We will discuss the algebraic meaning of \eqref{eq:TL}--\eqref{eq:R_tri} in Section~\ref{sec:spin_discussion}.
Thus we obtain the deformed HS chain, which is still chiral and of the form \eqref{eq:hamiltonian}. Further letting $\eta\to0$, both reduce to the isotropic HS hamiltonian $H^\textsc{hs}$, which is also obtained from $H^\text{Ino}$ as $\kappa \to 0$.

Finally, our model is integrable in the sense that the chiral hamiltonians \eqref{eq:hamiltonian} commute,
\begin{equation} \label{eq:comm_hams}
	[H^\textsc{l}, H^\textsc{r}] = 0 \, ,
\end{equation}
belonging to a tower of conserved charges whose expressions parallel those in \cite{lamers2022spin,MZ_23b}, see~\cite{KL_extended}.

\paragraph{Further properties.} The ordinary Inozemtsev chain has full $\mathit{SU}(2)$ spin symmetry. Our chain is its generalisation with spin symmetry broken to $\textit{U}(1)$: our conserved charges all commute with 
$S^z = \sum_i \sigma^z_{\!i}/2$.

Like the deformed HS chain, the spin interactions \eqref{eq:S^LR_diagr} involve multi\-spin interactions affecting all intermediate spins, whence the subscript `$[i,j]$'. While $\eta\neq 0$ breaks periodicity, our chain has quasiperiodic boundary conditions. One of the conserved charges is the deformed (lattice) translation operator (cf.~\cite{Lam_18}) 
\begin{gather}
	\vphantom{h} \nonumber \\[.5\baselineskip]
	\label{eq:transl} 
	G = \!
	\smash{
	\tikz[baseline={([yshift=-.5*11pt*0.2]current bounding box.center)},xscale=0.4,yscale=0.2,font=\footnotesize]{
		\node at (5,7-.1) {$\tikz[baseline={([yshift=-.5*11pt*.25]current bounding box.center)},scale=.35]{\fill[black, yshift=-.2] (0,0) rectangle ++(.4,.4)}$};
		\draw[rounded corners=2pt,->] (0,0) node[below]{$1$} -- (0,1) -- (5,6) -- (5,8.5) node[above]{$1$};
		\draw[rounded corners=2pt,->] (1,0) node[below]{$2$} -- (1,1) -- (0,2) -- (0,8.5) node[above]{$2$};
		\foreach \x in {2,...,4} \draw[rounded corners=2pt,->] (\x,0) -- (\x,\x) -- (\x-1,\x+1) -- (\x-1,8.5);
		\foreach \x in {-1,...,1} \draw (3+.2*\x,-1.4) node{$\cdot\mathstrut$};
		\foreach \x in {-1,...,1} \draw (2+.2*\x,9.4) node{$\cdot\mathstrut$};
		\draw[rounded corners=2pt,->] (5,0) node[below]{$N$} -- (5,5) -- (4,6) -- (4,8.5) node[above]{$N$};
		\node at (-.5,4.25) {$a$};
	} } \!\! = \, K_{\mspace{-2mu}N} \, P_{N-1\mspace{-1mu},\mspace{1mu}N}(1-N) \cdots P_{12}(-1) \, , \quad K_{\mspace{-2mu}N} = \E^{-\kappa\mspace{1mu}\eta\mspace{2mu}[a-(\sigma^z_1 + \dots + \sigma^z_{N-1})]\mspace{2mu}\sigma^z_N} \, .
	\\[.5\baselineskip] \vphantom{h} \nonumber
\end{gather}
Here $K_N$ is a diagonal twist, $\E^{-\kappa\mspace{1mu}\eta\mspace{1mu}a\mspace{1mu}\sigma^z} =  \text{diag}(\E^{-\kappa\mspace{1mu}\eta\mspace{1mu}a\mspace{1mu}},\E^{\kappa\mspace{1mu}\eta\mspace{1mu}a\mspace{1mu}})$, acting at site~$N$ with a shift of $a$ as in \eqref{eq:deformed_permutation}. 
Upon normalisation, \eqref{eq:transl} provides a notion of momentum, plus all $N$ eigenvectors at $S^z = N/2 - 1$ (cf.\ \textsection1.2.6 in \cite{lamers2022spin}), i.e.\ the magnons of our chain. 
Namely, $G^N = K_1 \cdots K_N$ is a central element, so the rescaled shift operator $G' = (K_1 \cdots K_N)^{-1/N} \, G$ obeys $G'^{\mspace{2mu}N} = 1$, and thus has eigenvalues $\E^{\I\,p}$ with `deformed momentum' $p$ quantised as $p = 2\pi \,n/N$ for $0\leqslant n < N$. The $N$ `deformed magnons' $\sum_{j=1}^N \E^{\I \mspace{2mu} p \mspace{1mu} j} G'^{\mspace{2mu}1-j} \, \ket{\downarrow\uparrow\cdots \uparrow}$ by construction have deformed momentum $p$. These are the simplest eigenstates after the reference vector $\ket{\uparrow\cdots \uparrow}$.
We have not yet been able to find a compact expression for 
their chiral dispersion relations. Moreover, \eqref{eq:transl} allows us to express the long-range interaction of neighbouring spins on sites $1$ and $N$ as
 \begin{equation} \label{eq:S_1N_via_G}
	S_{[1,N]}^{\textsc{l}} = G \, S_{[1,2]}^{\textsc{l}} \, G^{-1} , \quad 
	S_{[1,N]}^{\textsc{r}} = G^{-1} S_{[N-1,N]}^{\textsc{r}} \, G \, ,
\end{equation}
underlining the chirality of the hamiltonians~\eqref{eq:hamiltonian}.

\paragraph{New limits.} Our chain has various new limits. 
For $N\to\infty$ we formally get a hyperbolic counterpart of the deformed HS chain, with $N \leftrightsquigarrow \I\pi/\kappa$ and sum in \eqref{eq:hamiltonian} over all integers. Numerics suggests that its matrix entries converge. 

As discussed in the previous section, the limit $\eta \to 0$ yields the Inozemtsev spin chain (up to a conjugation).
Interestingly, this limit can be refined to obtain an intermediate spin chain that seems to be new, by setting $a= a'/\eta$ before sending $\eta\to0$. This does not affect the limits of the potential and deformed spin permutation, but changes the limit of the deformed exchange \eqref{eq:deformed_exchange} as a function of $a'$.
Both chiral hamiltonians \eqref{eq:hamiltonian} then limit to
\begin{equation} \label{eq:iso_dyn_ham}
	H^\text{Ino}(a') = \frac{1}{2} \sum_{i<j}^N \, \biggl( \phi'(i-j,a') \, \frac{\sigma^+_i \sigma^-_j}{2}
	+ \phi'(i-j,-a') \, \frac{\sigma^-_i \sigma^+_j}{2}
	+ \bar{V}^{\text{Ino}}(i-j) \, (1 - \sigma^z_i \sigma^z_j)
	\biggr) \, ,
\end{equation}
where $\phi'$ is the derivative with respect to the first variable of $\phi(x,y) = \theta(x+y)/[\theta(x) \theta(y)]$.
The hamiltonian \eqref{eq:iso_dyn_ham} generalises $H^\text{Ino}$ from \eqref{eq:iso} and \eqref{eq:pot_Ino} with an extra parameter $a'$ that breaks the left-right symmetry and $SU(2)$ spin symmetry. Unlike for $\eta \neq 0$, \eqref{eq:iso_dyn_ham} is not dynamical in the sense that the parameter $a'$ does not receive any shifts as in e.g.\ \eqref{eq:deformed_permutation}.
The spectrum is $a'$-dependent and real when $a' \in \I \, \R$. 
The isotropic Inozemtsev chain is retrieved by sending $a' \to 0$ or $a'\to \I \pi/\kappa$, since then $\phi'(x,a') \to \rho(x) = -\bar{V}^\text{Ino}(x)$. 

Finally, we turn to the short-range limit $\kappa \to \infty$. It is convenient to represent the potential \eqref{eq:pot} as the sum
\begin{equation} \label{eq:rho_diff_sum}
	\begin{aligned}
	\rho(x+\eta) - \rho(x-\eta) &= \sum_{n\in\mathbb{Z}} \frac{2 \mspace{2mu}\kappa\, \sinh(2\,\kappa\,\eta) }{ \sinh[\kappa\mspace{2mu}(\eta+x+N\mspace{2mu}n)]\,\sinh[\kappa\mspace{2mu}(\eta-x-N\mspace{2mu}n)]} \\
	&= 
	\sum_{n \in \mathbb{Z}} \frac{4\,\kappa \sinh(2\,\kappa\,\eta) }{\cosh(2\,\kappa\,\eta) - \cosh [2\,\kappa\,(N\mspace{2mu}n+x)]} 
	\, . 
	\end{aligned}
\end{equation} 
For a convergent but non-zero limit as $\kappa \to \infty$ we must also send $\eta \to 0$ with $\kappa \,\eta$ fixed so that $\cosh (2 \kappa \eta)$ becomes constant. Thus we set $\eta = -\I\pi\,\gamma/\kappa$ and rescale \eqref{eq:rho_diff_sum} by a prefactor behaving as $n_{\eta}(\kappa) \sim \E^{2\kappa}/[4\kappa \sinh(2\kappa \eta)]$ to obtain
\begin{equation}
	n_{-\I \pi\gamma/\kappa}(\kappa) \, \bigl( \rho(x-\I\pi\gamma/\kappa) - \rho(x+\I\pi\gamma/\kappa) \bigr) \; \to \; \delta_{x,1} + \delta_{x,N-1} \, , \quad \kappa\to\infty \, , \quad x\in \{1,\ldots, N-1\} \, .
\end{equation}
A choice of normalisation that fits with all other limits is to rescale the hamiltonians \eqref{eq:hamiltonian} by $n_{\eta}(\kappa) = \sinh^2 \mspace{-2mu}\kappa/[\kappa^2 \, \theta(2\eta)]$. This is why we choose denominator $\theta(2\eta)$ in the potential~\eqref{eq:pot_app} rather than the $2\eta$ from \cite{klabbers2022coordinate}; when $\eta\to 0$ the two have the same behaviour.
Therefore, as $\kappa\to\infty$, we get a nearest-neighbour chain
\begin{equation} \label{eq:Heis_lim}
	H^{\textsc{xxz}} = \sum_{i=1}^{N-1} \! S^{\textsc{h}}_{[i,i+1]} \ + S^{\textsc{h}}_{[1\mspace{-1mu},\mspace{1mu}N]} \, .
\end{equation}
Here, the exchange $S^{\textsc{h}}_{[i,i+1]} = E_{i,i+1}^{\textsc{h}}\bigl(a-(\sigma^z_1+\dots+\sigma^z_{i-1})\bigr)$ is defined like in \eqref{eq:deformed_permutation} in terms of a generalisation of \eqref{eq:TL}: 
\begin{equation} \label{eq:E_Heis}
	E^{\textsc{h}}(a) = 
	\begin{pmatrix}
		0 & \color{gray!80}{0} & \color{gray!80}{0} & \color{gray!80}{0} \\
		\color{gray!80}{0} & \!\hphantom{-}\frac{\sin[\pi \gamma(a-1)]}{\sin[\pi\gamma a]} & \!{-}\frac{\sin[\pi \gamma(a+1)]}{\sin[\pi\gamma a]} & \color{gray!80}{0} \\[5pt]
		\color{gray!80}{0} & \!{-}\frac{\sin[\pi \gamma(a-1)]}{\sin[\pi\gamma a]} & \!\hphantom{-}\frac{\sin[\pi \gamma(a+1)]}{\sin[\pi\gamma a]} & \color{gray!80}{0}\\[4pt]
		\color{gray!80}{0} & \color{gray!80}{0} & \color{gray!80}{0} & 0
	\end{pmatrix} 
	\, .
\end{equation}
Since the two expressions in \eqref{eq:S_1N_via_G} coincide, the boundary term in \eqref{eq:Heis_lim} admits two forms 
\begin{equation} \label{eq:S_1N_Heis}
	S_{[1\mspace{-1mu},\mspace{1mu}N]}^{\textsc{h}} = G^{\textsc{h}} \, S_{[1\mspace{-1mu},\mspace{1mu}2]}^{\textsc{h}} \, G^{\textsc{h}\,-1} = G^{\textsc{h}\,-1} S_{[N-1\mspace{-1mu},\mspace{1mu}N]}^{\textsc{h}} \, G^{\textsc{h}} \, ,
\end{equation} 
where \eqref{eq:transl} becomes 
$G^{\textsc{h}} = K_{\mspace{-2mu}N}^{\textsc{h}} \, P_{N-1\mspace{-1mu},\mspace{1mu}N}^{\mspace{1mu}\textsc{h}} \cdots P_{12}^{\mspace{1mu}\textsc{h}}$, with twist $\E^{\I\pi\mspace{1mu}\gamma\mspace{2mu} a\mspace{1mu}\sigma^z}$ and permutation built from  $\check{R}^{\textsc{h}}(a)= 1 - \E^{-\I\pi\mspace{1mu}\gamma} E^{\textsc{h}}(a)$  as in \eqref{eq:deformed_permutation}. Note that the arguments $x$ have completely disappeared. This \textit{R}-matrix also appeared in a slightly different form in \cite{arnaudon2000towards}, see (5.28) therein.

The short-range limit \eqref{eq:Heis_lim} is a `dynamical' variant of the Heisenberg \textsc{xxz} chain. It is no longer chiral, but remains quasiperiodic, since the twist in \eqref{eq:S_1N_Heis} prevents removing $a$. When $\gamma \to0$ we obtain, once more up to a conjugation that vanishes as $a\to-\I\mspace{2mu}\infty$, the usual periodic Heisenberg \textsc{xxx} chain (Fig.~\ref{fg:spin_limits}).

\subsection{Discussion} \label{sec:spin_discussion}

\paragraph{Form of spin interactions.}
The long-range interactions~\eqref{eq:S^LR_diagr} are very specific generalisations of $1-P_{ij}$. The need for such involved interactions is more clear for the 
deformed HS chain, so as to maintain the HS chain's integrability, enhanced spin symmetry, and extremely simple exact spectrum~\cite{Lam_18,lamers2022spin}. In turn generalising the deformed HS chain, our spin chain must have similar spin interactions. 

\paragraph{Choice of \textit{R}-matrix.}
The deformed HS chain already uses an \textit{R}-matrix in its deformed permutations, viz.~\eqref{eq:R_tri}. Its enhanced spin symmetry requires \cite{bernard1993yang,drinfel1986degenerate} $\check{R}^\text{tri}$ to be related (by `Baxterisation') to the Hecke algebra\,---\,and, for spin~1/2, the Temperley--Lieb algebra, see \eqref{eq:TL_rel} below. This necessarily leads to some asymmetry ($P \, \check{R} \, P \neq \check{R}$) as in \eqref{eq:TL}. Now, at the partially isotropic level, an elliptic potential asks for an \textit{R}-matrix with elliptic functions, cf.~\eqref{eq:deformed_exchange}. The standard choices are
\begin{itemize}
	\item Baxter's eight-vertex (\textsc{xyz}) \textit{R}-matrix \cite{Baxter1972193}: $P \, \check{R}^{\text{8v}} \, P = \check{R}^{\text{8v}}$, which generalises the symmetric six-vertex (\textsc{xxz}) \textit{R}-matrix; 
	\item Felder's elliptic dynamical \textit{R}-matrix~\eqref{eq:Rdyn} \cite{felder1994elliptic}: $S^z$-symmetric, which generalises the \textit{R}-matrix~\eqref{eq:R_tri}. 
\end{itemize} 
They are related by a (`face-vertex') transformation~\cite{baxter1973eight},
\begin{equation} \label{eq:FV}
	\check{R}^{\text{8v}}(x_i - x_j) \, T(x_i,x_j,a) = T(x_j,x_i,a) \, \check{R}(x_i - x_j,a) \, .
\end{equation}
One might expect the corresponding spin chains to be equivalent. Yet the resulting deformed exchanged interactions~\eqref{eq:deformed_exchange}, containing a derivative in $x$, are not related by the $x$-dependent transformation~\eqref{eq:FV}. It appears impossible to obtain \eqref{eq:R_tri} from $\check{R}^{\text{8v}}$ without \eqref{eq:FV}.%
\footnote{\ This is supported by the fact that the principal grading operator is essential in the construction of the universal elliptic \textit{R}-matrix of vertex type \cite{jimbo1999quasi}. We thank H.~Konno for pointing this out.} 
Hence our spin chain \emph{differs} from the (fully) anisotropic chain recently found by Matushko and Zotov using $\check{R}^\text{8v}$ \cite{MZ_23a}, which belongs to a landscape disjoint from Fig.~\ref{fg:spin_limits} \cite{KL_extended}. See \cite{klabbers2024landscapes} for a detailed analysis of this fact. 

\paragraph{Modular family.} 
As we will see below, `freezing' in fact produces an $\mathit{SL}(2,\mathbb{Z})$-family of integrable longe-range spin chains. Only two of these have a real spectrum for some parameter range, of which only \eqref{eq:hamiltonian} has a short-range limit. At the isotropic level this choice corresponds to shifting $\wp(x)$ to $-\rho'(x)$ \cite{Inozemtsev:1989yq,Inozemtsev_1995}; this shift also simplifies the dispersion and Bethe equations, and allows the latter to be recast in rational form \cite{klabbers2022coordinate}.

\paragraph{Algebraic structure at short range.} 
The operators $e_i \equiv S^\textsc{h}_{[i,i+1]}$ in \eqref{eq:Heis_lim} obey the Temperley--Lieb (TL) relations 
\begin{equation} \label{eq:TL_rel}
	e_i^2 = 2\cos(\pi\gamma)\,e_i \, , \quad 1\leqslant i \leqslant N-1 \, , \qquad 
	e_i \, e_{i\pm1} \, e_i = e_i \, , \quad 1\leqslant i \leqslant N-2 \, .
\end{equation}
The boundary term \eqref{eq:S_1N_Heis} is a `braid translation' \cite{martin1993algebraic}, and $e_0 \equiv S_{[1,N]}^{\textsc{h}}$ obeys the \emph{periodic} TL relations, i.e.\ the preceding extended to subscripts $\mathrm{mod}~N$. The translation $u \propto G^\textsc{h}$ enhances this to the \emph{affine} TL algebra, 
\begin{equation}
	u \, e_i \, u^{-1} = e_{i-1 \, \mathrm{mod} \, N} \, , \quad 1\leqslant i \leqslant N \, , \qquad 
	u^N \ \text{is central} \, , \qquad 
	u^2 \, e_1 \cdots e_{N-1} = e_{N-1} \, .
\end{equation}
Thus, \eqref{eq:Heis_lim} is a dynamical alternative to the twisted Heisenberg chain of~\cite{belletete2017correspondence}, relating to the affine TL algebra in a similar way as the usual TL algebra underpins the Heisenberg \textsc{xxz} chain with special open boundaries~\cite{pasquier1990common}. Also note that \eqref{eq:Heis_lim} resembles an unrestricted version of the \textsc{rsos} model~\cite{andrews1984eight}. It provides an $S^z$-symmetric alternative to the TL representation from the conclusion of \cite{martin1994blob}, enabled by the dynamical nature of our $e_i$, cf.~\cite{filali2011spin}.

\section{The quantum many body system}
\label{sec:qmbs}

\subsection{Hamiltonians}
Now consider $N$ spin-$\frac{1}{2}$ particles with coordinates~$x_j$ moving on a circle. Given another parameter $\epsilon$, consider the shift operator 
\begin{equation} \label{eq:shift}
	\Gamma_i = \exp\bigl(- \I\, \hbar \, \epsilon \, \partial_{x_i}\bigr) \, , \quad x_k \mapsto x_k - \I \, \hbar\, \epsilon \, \delta_{jk} \, .
\end{equation}
Our QMBS is given by a tower of conserved charges that are difference operators built from \eqref{eq:shift} and the deformed permutation \eqref{eq:deformed_permutation}. The first conserved charge is
\begin{align} \label{eq:Dtilde_1}
	\widetilde{D}_1 = {} & \sum_{i=1}^N A_i(\vect{x}) \times\! \tikz[baseline={([yshift=-.5*11pt*.2-1pt]current bounding box.center)},xscale=.4,yscale=.2,font=\footnotesize]{
		\draw[->] (10.5,0) node[below]{$x_N$} -- (10.5,10) node[above]{$x_N$};
		\draw[->] (9,0) -- (9,10);
		\foreach \x in {-1,...,1} \draw (9.25+.2*\x,-1) node{$\cdot\mathstrut$};
		\foreach \x in {-1,...,1} \draw (9.25+.2*\x,11.4) node{$\cdot\mathstrut$};
		\draw[very thick, rounded corners=2pt] (8,0) 	
			node[below]{$x^{\smash{\vect{-}}}_i$}
			-- (8,1) -- (5,4) -- (5,5) node{$\mathllap{\epsilon\,}\tikz[baseline={([yshift=-.5*11pt*.25]current bounding box.center)},scale=.35]{\fill[black] (0,0) circle (.2)}$};
		\draw[rounded corners=2pt,->] (5,5) -- (5,6) -- (8,9) -- (8,10) node[above]{$\smash{x_i}\vphantom{x_i}$};
		\draw[rounded corners=2pt,->] (7,0) -- (7,1) -- (8,2) -- (8,5) -- (8,8) -- (7,9) -- (7,10);
		\draw[rounded corners=2pt,->] (6,0) -- (6,2) -- (7,3) -- (7,5) -- (7,7) -- (6,8) -- (6,10);
		\foreach \x in {-1,...,1} \draw (6.5+.2*\x,-1) node{$\cdot\mathstrut$};
		\foreach \x in {-1,...,1} \draw (6.5+.2*\x,11.4) node{$\cdot\mathstrut$};
		\draw[rounded corners=2pt,->] (5,0) node[below]{$x_1$} -- (5,3) -- (6,4) -- (6,5) -- (6,6) -- (5,7) -- (5,10) node[above]{$x_1$};
		\foreach \x in {-1,...,1} \draw (9.75+.2*\x,5) node{$\cdot\mathstrut$};		
		\node at (4.4,2.5) {$a$};
	} 
	\qquad\quad
	\tikz[baseline={([yshift=-.5*11pt*.2-2pt]current bounding box.center)},xscale=.4,yscale=.2,font=\footnotesize]{
		\fill[black] (0,5) node {$\mathllap{\epsilon\,}\tikz[baseline={([yshift=-.5*11pt*.35]current bounding box.center)},scale=.35]{\fill[black] (0,0) circle (.2)}$}; 
		\draw[very thick] (0,3.5) node[below] 
			{$x^{\smash{\vect{-}}}_i$}
			-- (0,5.13);
		\draw[->] (0,5.13) -- (0,6.75) node[above] {$x_i$};
	} \!\! = \Gamma_i \, , \qquad x^{\smash{\vect{-}}}_i \equiv x_i -\I\,\hbar\,\epsilon \\[-.5ex] 
	= {} & \sum_{i=1}^N \, A_i(\vect{x}) \; P_{i-1\mspace{-1mu},\mspace{1mu}i}(x_i - x_{i-1}) \cdots P_{12}(x_i - x_1) \; \Gamma_i \; P_{12}(x_1 - x_i) \cdots P_{i-1\mspace{-1mu},\mspace{1mu}i}(x_{i-1} - x_i) \nonumber \\ 
	= {} & \sum_{i=1}^N \, A_i(\vect{x}) \; P_{i-1\mspace{-1mu},\mspace{1mu}i}(x_i - x_{i-1}) \cdots P_{12}(x_i - x_1) \;
	P_{12}(x_1 - x^{\smash{\vect{-}}}_i) \cdots P_{i-1\mspace{-1mu},\mspace{1mu}i}(x_{i-1} - x^{\smash{\vect{-}}}_i) \; \Gamma_i \, , \nonumber
\end{align}
with coefficients
\begin{equation}
	A_i(\vect{x}) = \prod_{j (\neq i)}^N \frac{\theta(x_i - x_j + \eta)}{\theta(x_i - x_j)} \, .
\end{equation}
We furthermore have an `anti\-chiral' version of \eqref{eq:Dtilde_1},
\begin{align} \label{eq:Dtilde_-1}
	\widetilde{D}_{-1} & = \sum_{i=1}^N \, A_i(-\vect{x}) \times \!\! \tikz[baseline={([yshift=-.5*11pt*.15-2pt]current bounding box.center)},xscale=-.4,yscale=.2,font=\footnotesize]{
		\draw[->] (10.5,0) node[below]{$x_1$} -- (10.5,10) node[above]{$\smash{x_1}$};
		\draw[->] (9,0) -- (9,10);
		\foreach \x in {-1,...,1} \draw (9.25+.2*\x,-1) node{$\cdot\mathstrut$};
		\foreach \x in {-1,...,1} \draw (9.25+.2*\x,11.4) node{$\cdot\mathstrut$};
		\draw[very thick,rounded corners=2pt] (8,0) 
			node[below]{$x^{\smash{\scriptsize\vect{+}}}_i\!\!$} 
			-- (8,1) -- (5,4) -- (5,5) node{$\tikz[baseline={([yshift=-.5*12pt*.35]current bounding box.center)},scale=.35]{\fill[black] (0,0) circle (.2)}\mathrlap{\,{-}\epsilon}$};
		\draw[rounded corners=2pt,->] (5,5) -- (5,6) -- (8,9) -- (8,10) node[above]{$\smash{x_i}$};
		\draw[rounded corners=2pt,->] (7,0) -- (7,1) -- (8,2) -- (8,5) -- (8,8) -- (7,9) -- (7,10);
		\foreach \x in {-1,...,1} \draw (6.5+.2*\x,-1) node{$\cdot\mathstrut$};
		\foreach \x in {-1,...,1} \draw (6.5+.2*\x,11.4) node{$\cdot\mathstrut$};
		\draw[rounded corners=2pt,->] (6,0) -- (6,2) -- (7,3) -- (7,5) -- (7,7) -- (6,8) -- (6,10);
		\draw[rounded corners=2pt,->] (5,0) node[below]{$x_N$} -- (5,3) -- (6,4) -- (6,5) -- (6,6) -- (5,7) -- (5,10) node[above]{$\smash{x_N}$};
		\foreach \x in {-1,...,1} \draw (9.75+.2*\x,5) node{$\cdot\mathstrut$};		
		\node at (11.1,2.5) {$a$};
		} 
	\qquad\qquad
	x^{\smash{\scriptsize\vect{+}}}_i \equiv x_i + \I\,\hbar\,\epsilon 
	\\[-.5ex]
	& = \sum_{i=1}^N \, A_i(-\vect{x}) \, P_{i,i+1}(x_{i+1} - x_i) \cdots P_{N-1\mspace{-1mu},\mspace{1mu}N}(x_N - x_i) \; \Gamma_i^{-1} \; P_{N-1\mspace{-1mu},\mspace{1mu}N}(x_i - x_N) \cdots P_{i,i+1}(x_i - x_{i+1}) \nonumber \\
	& = \sum_{i=1}^N \, A_i(-\vect{x}) \, P_{i,i+1}(x_{i+1} - x_i) \cdots P_{N-1\mspace{-1mu},\mspace{1mu}N}(x_N - x_i) \; 
	P_{N-1\mspace{-1mu},\mspace{1mu}N}(x_i^{\smash{\vect{+}}} - x_N) \cdots P_{i,i+1}(x_i^{\smash{\vect{+}}} - x_{i+1}) \; \Gamma_i^{-1} \, . \nonumber
\end{align}
These two operators commute with each other, and with the total shift operator 
\begin{equation} \label{eq:Dtilde_N}
	\widetilde{D}_N = \Gamma_1 \cdots \Gamma_N \, .
\end{equation}
In Section~\ref{sec:heuristics} we will describe how the higher conserved charges, whose structure is like in \cite{Che_94a,lamers2022spin,MZ_23a}, are constructed.

\paragraph{Example.} For $N=3$ we have
\begin{gather}
	\begin{aligned}
		\widetilde{D}_1 = {} & A_1(\vect{x}) \, \Gamma_1 \; + \; A_2(\vect{x}) \, P_{12}(x_2 - x_1) \, \Gamma_2 \, P_{12}(x_1 - x_2) \\
		& +  A_3(\vect{x}) \, P_{23}(x_3 - x_2) \, P_{12}(x_3 - x_1) \, \Gamma_3 \, P_{12}(x_1 - x_3) \, P_{23}(x_2 - x_3) \, , \\
		\widetilde{D}_{-1} = {} & A_3(-\vect{x}) \, \Gamma_3^{-1} \; + \; A_2(-\vect{x}) \, P_{23}(x_3 - x_2) \, \Gamma_2^{-1} \, P_{23}(x_2 - x_3) \\
		& +  A_1(-\vect{x}) \, P_{12}(x_2 - x_1) \, P_{23}(x_3 - x_1) \, \Gamma_1^{-1} \, P_{23}(x_1 - x_3) \, P_{12}(x_1 - x_2) \, .
	\end{aligned}
\end{gather}

\subsection{Properties and limits}

Our QMBS, of which \eqref{eq:Dtilde_1} and \eqref{eq:Dtilde_-1} are the first two commuting charges,  depends on the four parameters of our spin chain, as well as on the shift~$\epsilon$.

\paragraph{Defining properties.} 
As $\eta \to 0$, again with $a\to -\I\,\infty$, we get the (`effective' form of the) elliptic spin-CS system \cite{hikami1993integrability, krichever1994spin}. Next, $\kappa\to 0$ and $a\to -\I\,\infty$ readily yields the spin-RM system \cite{Che_94a,lamers2022spin} underlying the deformed HS chain \cite{bernard1993yang, Ugl_95u, lamers2022spin}. See Fig.~\ref{fg:qmbs_limits}. 
Replacing $P(x) \rightsquigarrow 1$ gives the spinless elliptic Ruijsenaars system~\cite{Rui_87}. 

Moreover, our QMBS is integrable in the sense that the difference operators all commute, e.g.\
\begin{equation} \label{eq:Dtilde_commutativity}
	[\widetilde{D}_1,\widetilde{D}_{-1}] = 0 \, , \quad [\widetilde{D}_{\pm1},\widetilde{D}_N] = 0 \, .
\end{equation}
The second equality is clear as $\widetilde{D}_{\pm1}$ only depend on coordinate differences. The first one can be checked explicitly for low $N$.

\subsection{Discussion}

\paragraph{Commutativity.} 
It seems difficult to use the proof of integrability of \cite{MZ_23a}, which relies heavily on the periodicity properties of $\check{R}^{\text{8v}}$ for simplifying expressions and setting up a proof by induction. Alas, \eqref{eq:Rdyn} does not have such simple properties. Our proof of \eqref{eq:Dtilde_commutativity} is independent. In view of its technical nature it will appear elsewhere.

\paragraph{Choice of \textit{R}-matrix.}
Since \eqref{eq:Dtilde_1}--\eqref{eq:Dtilde_-1} only differ from the spin-Ruijsenaars model found by Matushko and Zotov~\cite{MZ_23a} in the choice of \textit{R}-matrix,
\eqref{eq:FV} might again lead one to expect these QMBSs to be equivalent. But, because the face-vertex transformation \eqref{eq:FV} depends on coordinates~$x_k$, it does not commute with the shift operators~$\Gamma_i$. Thus our difference operators are \emph{not} face-vertex transforms of those of MZ, and define \emph{another} QMBS. As we have seen, this difference persists to all limiting spin chains (see \cite{klabbers2024landscapes} for more). 

\paragraph{Modular family.}
A new feature of the elliptic case is that there is an $\mathit{SL}(2,\mathbb{Z})$-family of classical equilibria of \eqref{eq:freezing_stationary} related by modular transformations of the quasiperiods $N,\I \pi/\kappa$ \cite{KL_extended}. These equilibria can be identified by repara\-me\-trising $\eta,a,\epsilon,\vect{x}$. Upon freezing, however, each equilibrium yields a \emph{different} integrable spin chain.

\subsection{Heuristic derivation of the QMBS}
\label{sec:heuristics}

Let us motivate how the expressions \eqref{eq:Dtilde_1}, \eqref{eq:Dtilde_-1} and \eqref{eq:Dtilde_N} for the charges of our QMBS with spins 
can be `derived' from the \emph{spinless} QMBS known as the elliptic Ruijsenaars system. The latter describes $N$ scalar particles moving on a circle with coordinates $x_k$ and is defined by the difference operator 
\begin{equation} \label{eq:Ruij}
	D_1 = \sum_{i=1}^N \, A_i(\vect{x}) \, \Gamma_i \, , \qquad A_i(\vect{x}) = \! \prod_{j (\neq i)}^N \!\! \frac{\theta(x_i - x_j + \eta)}{\theta(x_i - x_j)} \, . 
\end{equation}
The operator $D_1$ belongs to a hierarchy of conserved charges, i.e.\ commuting difference operators. While this commutativity holds in general, it is physically reasonable to focus on bosonic/fermionic wave functions with definite (anti)symmetry
\begin{equation} \label{eq:bosons/fermions_spinless}
	s_{i,i+1} \, \Psi(\vect{x}) = \pm\Psi(\vect{x}) \, , \qquad 1 \leqslant i<N \, . 
\end{equation}
The space of either type of wave functions is preserved by \eqref{eq:Ruij}. At the same time, on either space, \eqref{eq:Ruij} is determined by any single term: if we have an operator of the form $\sum_i B_i(\vect{x}) \, \Gamma_i$ where, say, $B_1(\vect{x}) = A_1(\vect{x})$ is as in \eqref{eq:Ruij}, then the prescribed symmetry fixes the remaining coefficients to be as in \eqref{eq:Ruij} too. Indeed, on any wave function obeying \eqref{eq:bosons/fermions_spinless} we have $D_1 \Psi(\vect{x}) = (\pm s_{12}) \, D_1 \, (\pm s_{12}) \, \Psi(\vect{x}) = s_{12} \, D_1 \, s_{12} \, \Psi(\vect{x})$ since $D_1 \Psi(\vect{x})$ also obeys \eqref{eq:bosons/fermions_spinless}; comparing coefficients of $\Gamma_2$ in $D_1 = s_{12} \, D_1 \, s_{12}$ gives $B_2(\vect{x}) = s_{12} \, B_1(\vect{x}) \, s_{12} = A_2(\vect{x})$. Likewise, equating coefficients of $\Gamma_3$ in $D_1 = s_{23} \, D_1 \, s_{23}$ yields $B_3(\vect{x}) = A_3(\vect{x})$, and so on. This argument provides a useful heuristic to understand the structure of Ruijsenaars operators in more complicated settings, such as the trigonometric spin-Ruijsenaars--Macdonald system~\cite{lamers2022spin}, the trigonometric and elliptic spin-Ruijsenaars systems of Matushko and Zotov \cite{MZ_23a}, and ours.

Now consider a QMBS with $N$ \emph{spin}-1/2 particles moving on a circle. To define bosons or fermions in our setting, the appropriate permutation operator for the particles is 
\begin{equation}
	P_{i,i+1}^\text{tot} = s_{i,i+1} \, P_{i,i+1}(x_i - x_{i+1})\, , 
\end{equation}
which exchanges both coordinates, through $s_{i,i+1}$, as well as spins, through $P_{i,i+1}(x_i-x_{i+1})$ as defined in \eqref{eq:deformed_permutation}. Such permutation operators form a representation of the braid group (see Appendix \ref{app:def_perm}), and reduce to the usual permutation of particles, $s_{i,i+1} P_{i,i+1}$, as $\eta \to 0$. In terms of this permutation operator  the boson(fermion) condition is simply
\begin{equation} \label{eq:bosons/fermions_spin}
	P_{i,i+1}^\text{tot} \, \ket{\Psi} 
	= \pm \ket{\Psi} \, , \qquad 1 \leqslant i<N \, . 
\end{equation}
Now suppose a difference operator has the form $\widetilde{D}_1 = \sum_i \widetilde{B}_i(\vect{x}) \, \Gamma_i$ on either space, and again $\widetilde{B}_1(\vect{x}) = A_1(\vect{x})$. The coefficient of $\Gamma_2$ in $\widetilde{D}_1 = P_{12}^\text{tot} \, \widetilde{D}_1 \, P_{12}^\text{tot}$ can be found by comparing
\begin{equation}
	\begin{aligned}
		\widetilde{B}_2(\vect{x}) \, \Gamma_2 = P_{12}^\text{tot} \, \widetilde{B}_1(\vect{x}) \, \, \Gamma_1 \, P_{12}^\text{tot}  & = s_{12} \, P_{12}(x_1 - x_2) \, A_1(\vect{x}) \, \Gamma_1 \, s_{12} \, P_{12}(x_1 - x_2) \\ 
		& = A_2(\vect{x}) \, P_{12}(x_2 - x_1) \, \Gamma_2 \, P_{12}(x_1 - x_2) \\ 
		& = A_2(\vect{x}) \, P_{12}(x_2 - x_1) \, P_{12}(x_1 - x_2 + \I \, \hbar \, \epsilon) \, \Gamma_2 \, ,
	\end{aligned}
\end{equation}
whence $\widetilde{B}_2(\vect{x}) = A_2(\vect{x}) \, P_{12}(x_2 - x_1) \, P_{12}(x_1 - x_2 + \I \, \hbar \,\epsilon)$. Similarly,
\begin{equation}
	\begin{aligned}
		\widetilde{B}_3(\vect{x}) \, \Gamma_3 &= P_{23}^\text{tot} \, \widetilde{B}_2(\vect{x}) \, \, \Gamma_2 \, P_{23}^\text{tot} \\
		& = s_{23} \, P_{23}(x_2 - x_3) \, A_2(\vect{x}) \, P_{12}(x_2 - x_1) \, \Gamma_2 \, P_{12}(x_1 - x_2) \, s_{23} \, P_{23}(x_2 - x_3) \\ 
		& = A_3(\vect{x}) \, P_{23}(x_3 - x_2) \, P_{12}(x_3 - x_1) \, \Gamma_3 \,P_{12}(x_1 - x_3) \, P_{23}(x_2 - x_3) \\ 
		& = A_3(\vect{x}) \, P_{23}(x_3 - x_2) \, P_{12}(x_3 - x_1) \, P_{12}(x_1 - x_3 + \I \, \hbar \, \epsilon) \, P_{23}(x_2 - x_3 + \I \, \hbar \, \epsilon) \, \Gamma_3 \, ,
	\end{aligned}
\end{equation}
and so on. In this way we obtain our first difference operator \eqref{eq:Dtilde_1}. 

Its `antichiral' counterpart $\widetilde{D}_{-1} = \sum_i \widetilde{B}_{-i}(\vect{x}) \, \Gamma_i^{-1}$ is likewise fixed by \eqref{eq:bosons/fermions_spin} starting from the coefficient $\widetilde{B}_{-N}(\vect{x}) = A_N(-\vect{x})$ and yields \eqref{eq:Dtilde_-1}. 

More generally, the higher conserved charges $\widetilde{D}_{\pm r} = \sum_{i_1<\dots<i_r}^N \widetilde{B}_{\pm i_1,\dots,\pm i_r}(\vect{x}) \, \Gamma_{i_1}^{\pm 1} \cdots \Gamma_{i_r}^{\pm 1}$ are obtained in the same way from $\widetilde{B}_{1\dots r}(\vect{x}) = A_{1\dots r}(\vect{x}) = \prod_{i(\leqslant r)} \prod_{j(>r)}^N \theta(x_i - x_j + \eta)/\theta(x_i - x_j)$ and $\widetilde{B}_{-(N-r+1),\dots,-N}(\vect{x}) = A_{N-r+1,\dots N}(-\vect{x})$, yielding a tower of hamiltonians, with structure like in \cite{Che_94a,lamers2022spin,MZ_23a}. In particular, the total shift operator takes the simple form $\widetilde{D}_N = \Gamma_1 \cdots \Gamma_N$. 

We emphasise that while this argument `explains' the structure of our dynamical spin-Ruijsenaars operators, including the appearance of \textit{R}-matrices, and shows that our operators preserve the `physical space' of bosonic/fermionic vectors~\eqref{eq:bosons/fermions_spin}, it does \emph{not} prove their commutativity \eqref{eq:Dtilde_commutativity}. 
The proof will be published elsewhere in view of its technical nature.

\subsection{Freezing} \label{sec:freezing}

Let us discuss the relation between the spin-chain hamiltonians \eqref{eq:hamiltonian} and the spin-Ruijsenaars operators \eqref{eq:Dtilde_1}--\eqref{eq:Dtilde_-1}. We begin with a useful heuristics for deriving the spin-chain hamiltonians from the \textsc{qmbs}. 
Let $\delta = \partial_{\epsilon}\big|_{\epsilon=0}$ 
denote linearisation in $\epsilon$. Using 
$\delta\mspace{1mu}\Gamma_j = -\I \, \hbar\, \partial_{x_j}$
and the Leibniz rule we compute
\vspace{-0.3cm}
\begin{equation}
	\begin{aligned}
	\delta \widetilde{D}_1 
	& = \sum_{j=1}^N A_j(\vect{x}) \times \delta \, \tikz[baseline={([yshift=-.5*11pt*.2]current bounding box.center)},xscale=.4,yscale=.2,font=\footnotesize]{
		\draw[->] (10.5,0) node[below]{$x_N$} -- (10.5,10) node[above]{$x_N$};
		\draw[->] (9,0) -- (9,10);
		\foreach \x in {-1,...,1} \draw (9.25+.2*\x,-1) node{$\cdot\mathstrut$};
		\foreach \x in {-1,...,1} \draw (9.25+.2*\x,11.4) node{$\cdot\mathstrut$};
		\draw[very thick, rounded corners=2pt] (8,0) 
			node[below]{$x^{\smash{\vect{-}}}_j$} 
			-- (8,1) -- (5,4) -- (5,5) node{$\mathllap{\epsilon\,}\tikz[baseline={([yshift=-.5*11pt*.25]current bounding box.center)},scale=.35]{\fill[black] (0,0) circle (.2)}$};
		\draw[rounded corners=2pt,->] (5,5) -- (5,6) -- (8,9) -- (8,10) node[above]{$\smash{x_j}\vphantom{x_i}$};
		\draw[rounded corners=2pt,->] (7,0) -- (7,1) -- (8,2) -- (8,8) -- (7,9) -- (7,10);
		\draw[rounded corners=2pt,->] (6,0) -- (6,2) -- (7,3) -- (7,7) -- (6,8) -- (6,10);
		\foreach \x in {-1,...,1} \draw (6.5+.2*\x,-1) node{$\cdot\mathstrut$};
		\foreach \x in {-1,...,1} \draw (6.5+.2*\x,11.4) node{$\cdot\mathstrut$};
		\draw[rounded corners=2pt,->] (5,0) node[below]{$x_1$} -- (5,3) -- (6,4) -- (6,5) -- (6,6) -- (5,7) -- (5,10) node[above]{$x_1$};
		\foreach \x in {-1,...,1} \draw (9.75+.2*\x,5) node{$\cdot\mathstrut$};		
		\node at (4.4,2.5) {$a$};
	} \\[1.5ex]
	& = 
	\sum_{j=1}^N A_j(\vect{x}) \times 
	-\I\,\hbar 
	\, \left(
	\partial_{x_j} - \sum_{i=1}^{j-1}
	\vphantom{\tikz[baseline={([yshift=-.5*11pt*.2*1.2]current bounding box.center)},yscale=.13*1.2]{\draw[->] (0,0) -- (0,10);} }
	\smash{
	\tikz[baseline={([yshift=-.5*11pt*.2]current bounding box.center)},xscale=.4,yscale=.2,font=\footnotesize]{
		\draw[->] (10.5,0) node[below]{$x_N$} -- (10.5,10) node[above]{$x_N$};
		\draw[->] (9,0) -- (9,10);
		\foreach \x in {-1,...,1} \draw (9.25+.2*\x,-1) node{$\cdot\mathstrut$};
		\foreach \x in {-1,...,1} \draw (9.25+.2*\x,11.4) node{$\cdot\mathstrut$};
		\draw[rounded corners=2pt,->] (8,0) node[below]{$x_j$} -- (8,1) -- (5,4) -- (5,6) -- (8,9) -- (8,10) node[above]{$\smash{x_j}\vphantom{x_i}$};
		\draw[rounded corners=2pt,->] (7,0) -- (7,1) -- (8,2) -- (8,8) -- (7,9) -- (7,10);
		\draw[rounded corners=2pt,->] (6,0) node[below] {$x_i$} -- (6,2) -- (7,3) -- (7,7) -- (6,8) -- (6,10) node[above] {$x_i$};
		\node at (6.5,2.5cm-.5pt) {\rotatebox{30}{$\circledast$}};
		\draw[rounded corners=2pt,->] (5,0) -- (5,3) -- (6,4) -- (6,6) -- (5,7) -- (5,10);
		\foreach \x in {-1,...,1} \draw (9.75+.2*\x,5) node{$\cdot\mathstrut$};		
		\node at (4.4,3.5) {$a$};
	} } \right) \, , \\
	\vspace{1.5cm}
	\end{aligned}
\end{equation}
where the $\circledast$ denotes a derivative of the (deformed) permutation \eqref{eq:deformed_permutation}. 
Note that the spin and differential part decouple (`spin-charge separation'). By unitarity and recognising \eqref{eq:deformed_exchange} the spin part is 
\begin{equation}
	\begin{aligned}
	\tikz[baseline={([yshift=-.5*11pt*.2-1pt]current bounding box.center)},xscale=.4,yscale=.2,font=\footnotesize]{
		\draw[->] (10.5,0) node[below]{$x_N$} -- (10.5,10) node[above]{$x_N$};
		\draw[->] (9,0) -- (9,10);
		\foreach \x in {-1,...,1} \draw (9.25+.2*\x,-1) node{$\cdot\mathstrut$};
		\foreach \x in {-1,...,1} \draw (9.25+.2*\x,11.4) node{$\cdot\mathstrut$};
		\draw[rounded corners=2pt,->] (8,0) node[below]{$x_j$} -- (8,1) -- (5,4) -- (5,5) -- (5,6) -- (8,9) -- (8,10) node[above]{$\smash{x_j}\vphantom{x_i}$};
		\draw[rounded corners=2pt,->] (7,0) -- (7,1) -- (8,2) -- (8,5) -- (8,8) -- (7,9) -- (7,10);
		\draw[rounded corners=2pt,->] (6,0) node[below] {$x_i$} -- (6,2) -- (7,3) -- (7,5) -- (7,7) -- (6,8) -- (6,10) node[above] {$x_i$};
		\node at (6.5,2.5cm-.5pt) {\rotatebox{30}{$\circledast$}};
		\draw[rounded corners=2pt,->] (5,0) -- (5,3) -- (6,4) -- (6,5) -- (6,6) -- (5,7) -- (5,10);
		\foreach \x in {-1,...,1} \draw (9.75+.2*\x,5) node{$\cdot\mathstrut$};		
		\node at (4.3,5) {$a$};
	} \! & = \,
		\tikz[baseline={([yshift=-.5*11pt*.2-2pt]current bounding box.center)},xscale=.4,yscale=.2,font=\footnotesize]{
		\draw[->] (10.5,1) node[below]{$x_N$} -- (10.5,9) node[above]{$x_N$};
		\draw[->] (9,1) -- (9,9);
		\foreach \x in {-1,...,1} \draw (9.25+.2*\x,0) node{$\cdot\mathstrut$};
		\foreach \x in {-1,...,1} \draw (9.25+.2*\x,10.4) node{$\cdot\mathstrut$};
		\draw[rounded corners=2pt,->] (8,1) node[below]{$x_j$} -- (8,2) -- (6,4) -- (6,6) -- (8,8) -- (8,9) node[above]{$\smash{x_j}\vphantom{x_i}$};
		\draw[rounded corners=2pt,->] (7,1) -- (7,2) -- (8,3) -- (8,7) -- (7,8) -- (7,9);
		\draw[rounded corners=2pt,->] (6,1) node[below] {$x_i$} -- (6,3) -- (7,4) -- (7,6) -- (6,7) -- (6,9) node[above] {$x_i$};
		\node at (6.5,3.5cm-.5pt) {\rotatebox{30}{$\circledast$}};
		\draw[->] (5,1) -- (5,9);
		\foreach \x in {-1,...,1} \draw (9.75+.2*\x,5) node{$\cdot\mathstrut$};		
		\node at (4.3,5) {$a$};
	} \\[-.5mm]
	& = \, \theta(\eta) \, V(x_i - x_j) \times 
	\tikz[baseline={([yshift=-.5*11pt*.2-2pt]current bounding box.center)},xscale=.4,yscale=.2,font=\footnotesize]{
	\draw[->] (10.5,2) node[below]{$x_N$} -- (10.5,8) node[above]{$x_N$};
	\draw[->] (9,2) -- (9,8);
	\foreach \x in {-1,...,1} \draw (9.25+.2*\x,1) node{$\cdot\mathstrut$};
	\foreach \x in {-1,...,1} \draw (9.25+.2*\x,9.4) node{$\cdot\mathstrut$};
	\draw[rounded corners=2pt,->] (8,2) node[below]{$x_j$} -- (8,3) -- (7,4) -- (7,6) -- (8,7) -- (8,8) node[above]{$\smash{x_j}\vphantom{x_i}$};
	\draw[rounded corners=2pt,->] (7,2) -- (7,3) -- (8,4) -- (8,6) -- (7,7) -- (7,8);
	\draw[rounded corners=2pt,->] (6,2) node[below] {$x_i$} -- (6,8) node[above] {$x_i$};
	\draw[style={decorate, decoration={zigzag,amplitude=.5mm,segment length=1mm}}] (6,5) -- (7,5);
	\draw[->] (5,2) -- (5,8);
	\foreach \x in {-1,...,1} \draw (9.75+.2*\x,5) node{$\cdot\mathstrut$};		
	\node at (4.3,5) {$a$};
	} ,
	\end{aligned}
\end{equation}
which equals $\theta(\eta) \, V(i-j) \, S^\textsc{l}_{[i,j]}$ at $x_k^\star = k$ ($1\leqslant k \leqslant N$). 
The computation of $\delta\widetilde{D}_{-1}$ is analogous, instead yielding $\theta(\eta) \, V(i-j) \, S^\textsc{r}_{[i,j]}$. 
As we will explain below, at the equispaced positions $x_k^\star = k$ the coefficients $A_j(\vect{x}^\star) = A^\star$ have a common value [$A^\star = \theta(\eta)_{N=1}/N\,\theta(\eta)$]. Then we can conclude that
\begin{equation} \label{eq:freezing_app}
	 \begin{aligned} 
	 \frac{1}{\I \, \hbar \, \theta(\eta)} \, \biggl[ \delta\widetilde{D}_{\pm 1} \mp \sum_{j=1}^N A_j(\pm\vect{x}) \, \delta\,\Gamma_j \biggr]_{\raisebox{1mm}{\scriptsize $x_k = x_k^\star$}} &  = 
	 \frac{1}{\I \, \hbar \, \theta(\eta)} \, \biggl[ \delta\widetilde{D}_{ \pm 1} \, \mp \, A^\star \, \delta\widetilde{D}_N \biggr]_{\raisebox{1mm}{\scriptsize $x_k = x_k^\star$}} \\
	 & = 
	 A^\star \sum_{i<j}^N V(i-j) \, S^{\textsc{l},\mspace{1mu}\textsc{r}}_{[i,j]} = 
	 A^\star \, H^{\textsc{l},\mspace{1mu}\textsc{r}} \, .
	 \end{aligned}
\end{equation} 
The physical picture is that $\epsilon = \I \, \eta/g$ (cf.\ the `nonrelativistic limit' to the spin-Calogero--Sutherland system) and in the classical/strong-coupling limit $\hbar \, \epsilon \propto \hbar/g \to 0$ the kinetic energy is negligible compared to the potential energy, and the particles slow down to come to a halt, `freezing' at the classical equilibrium positions $x_k^\star = k$ of the spinless elliptic Ruijsenaars system. 

The expansion \eqref{eq:freezing_app} gives the correct spin-chain hamiltonian, but the calculation has to be made more precise to turn it into a proper derivation. Here we outline how this goes; details will be given in \cite{KL_extended}. Let us for a moment keep the elliptic parameter $\tau$ arbitrary by replacing the (odd) Jacobi theta function~\eqref{eq:theta} by 
\begin{equation}
	\vartheta(x\,|\,\tau) = \frac{\sin(\pi \, x)}{\pi} \prod_{n=1}^{\infty} \frac{ \sin [\pi(n \, \tau + x)] \sin [\pi( n \, \tau - x)] }{\sin^2 (\pi n \tau)} \, .
\end{equation}
Consider the classical spinless elliptic Ruijsenaars system with canonically conjugate coordinates $x_i$ and momenta $p_j$, with Poisson brackets $\{x_i, p_j\} = \delta_{ij}$. The (`chiral') hamiltonians are
\begin{equation} \label{eq:Ruij_class}
	D_{\pm1}^\text{cl} = \sum_{i=1}^N  \E^{\pm \epsilon \, p_i} \, A_i(\pm \vect{x};\eta\,|\,\tau) \, , \qquad A_i(\vect{x};\eta\,|\,\tau) = \! \prod_{j (\neq i)}^N \!\! \frac{\vartheta(x_i - x_j + \eta\,|\,\tau)}{\vartheta(x_i - x_j\,|\,\tau)} \, . 
\end{equation}
These functions belong to a family of $N$ independent Poisson-commuting quantities, which are the conserved charges of the classical Ruijsenaars--Schneider system~\cite{ruijsenaars1986new}. Picking $D_1^\text{cl}$ as hamiltonian defines a time flow with velocities
\begin{subequations} \label{eq:freezing_xj}
\begin{gather} 
	\frac{\partial x_j}{\partial t} \equiv \{x_j,D_1^\text{cl} \} = \frac{ \partial D_1^\text{cl} }{\partial p_j} = \, \epsilon \; \E^{\epsilon \, p_j} A_j(\vect{x};\eta\,|\,\tau) \, ,
\shortintertext{and momenta changing as}
\label{eq:freezing_pj}
	\frac{\partial p_j}{\partial t} \equiv \{p_j,D_1^\text{cl}\} = -\frac{\partial D_1^\text{cl}}{\partial x_j} = -\sum_{i=1}^N \E^{\epsilon \, p_i} \, \partial_{x_j} A_i(\vect{x};\eta\,|\,\tau)
	\,.  
\end{gather}
\end{subequations}
We can search for phase-space configurations $(\vect{x}^\star, \vect{p}^\star) \in \mathbb{C}^{2N}$ that satisfy the classical equilibrium conditions
\begin{equation} \label{eq:freezing_stationary}
	\frac{\partial x_j}{\partial t} = \epsilon \, A^\star\, , \quad \frac{\partial p_j}{\partial t} = 0\, , 
\end{equation}
for a ($j$-independent) constant $A^\star$. Such configurations are `frozen' in the sense that they remain stationary in the co-moving frame with velocity $A^\star$. Evaluating our quantum spin-Ruijsenaars system at such stationary configurations and dropping all derivatives in a consistent manner yields a spin-chain hamiltonian like in \eqref{eq:freezing_app}, cf.~\cite{MZ_23b}. 

One equilibrium configuration solving \eqref{eq:freezing_stationary} is 
\begin{equation} \label{eq:equilibr_1}
	x_j^\star = \frac{j}{N} \, , \quad p_j^\star = 0 \, , \quad \tau = \frac{\omega}{N} \, ,
\end{equation}
(we parametrise $\omega = \I \pi/\kappa$). In this case all coefficients $A_j\bigl(\vect{x}^\star ; \frac{\eta}{N} \,\big|\, \frac{\omega}{N}\bigr)$ 
are equal to the constant $A^\star \equiv \vartheta(\eta\,|\,\omega)/\bigl[N\,\vartheta\bigl(\frac{\eta}{N} \,\big|\, \frac{\omega}{N}\bigr)\bigr]$. This configuration is used to obtain an integrable spin chain by freezing for the HS and deformed HS chains \cite{lamers2022spin} and was used by Matushko and Zotov \cite{MZ_23b}. In this case the argument around \eqref{eq:freezing_app} can be made rigorous following \cite{MZ_23b}.

However, the resulting spin chain does not admit a Heisenberg-type short-range limit. Happily, there are many more solutions to \eqref{eq:freezing_stationary}, each belonging to a (lattice) parameter $\tau$ \cite{KL_extended}. The modular action of $\mathit{SL}(2,\mathbb{Z})$ on $\tau$ relates these solutions. In particular, one of the other equilibrium configurations is
\begin{equation} \label{eq:equilibr_2}
	x_j^\star = \frac{-j}{\omega} \, , \quad
	p_j^\star = \frac{\I\pi\mspace{2mu}\eta}{\omega\mspace{2mu}\epsilon} \, (N-2j+1) \, , \quad \tau^\star = \frac{{-}N}{\omega} \, ,
\end{equation} 
which yields the theta function \eqref{eq:theta} as $\theta(x) = \omega\, \vartheta\bigl(\frac{x}{\omega} \,\big|\, \frac{-N}{\omega}\bigr)$. Note that the positions in \eqref{eq:equilibr_2} are still equally spaced, albeit now along the imaginary axis. The values of the momenta in \eqref{eq:equilibr_2} compensate for the differences between 
\begin{equation}
	A_j\Bigl(\vect{x}^\star; \frac{{-}\eta}{\omega} \,\Big|\, \frac{-N}{\omega}\Bigr) = \E^{-(N-2j+1) \mspace{1mu} \eta \mspace{1mu} \kappa} \, \vartheta\Bigl(\frac{\eta}{\omega} \,\Big|\, \frac{-1}{\omega}\Bigr) / \vartheta\Bigl(\frac{\eta}{\omega} \,\Big|\, \frac{-N}{\omega}\Bigr)\, , 
\end{equation}
so that all velocities \eqref{eq:freezing_stationary} are again equal; one may think of the particles as having different masses. Thus, the expansion leading to \eqref{eq:freezing_app} has to be computed more carefully, taking into account that $\Gamma_i = \E^{\epsilon\,\hat{p}_i} \to \E^{\epsilon\,p_i}$ also contributes to the value of $A^\star = \vartheta\bigl(\frac{\eta}{\omega} \,\big|\, \frac{-1}{\omega}\bigr)/\vartheta\bigl(\frac{\eta}{\omega} \,\big|\, \frac{-N}{\omega}\bigr)$; see \cite{KL_extended} for details. The result is that freezing the quantum spin-Ruijsenaars system at \eqref{eq:equilibr_2} yields our spin-chain hamiltonians \eqref{eq:hamiltonian} with theta functions \eqref{eq:theta}.  Unlike the spin chain obtained by freezing at \eqref{eq:equilibr_1}, this spin chain admits a short-range limit, as discussed above.

Note that \eqref{eq:freezing_app} does not yet imply the commutativity \eqref{eq:comm_hams} of the commuting charges of our spin chain. This can be proven~\cite{KL_extended} following~\cite{TH_95,Ugl_95u,MZ_23b} using the commutativity \eqref{eq:Dtilde_commutativity} for the spin-Ruijsenaars system.
The conclusion is that the commutativity of the hamiltonians of our QMBS implies that for the hamiltonians of our spin chain.

\section{Conclusion} 
\label{sec:conclusion}

\paragraph{Summary.} We introduced a new integrable long-range quantum spin chain that unifies the Inozemtsev chain and the deformed Haldane--Shastry chain: the deformed Inozemtsev chain. 
It is obtained by `freezing' a quantum many body system (QMBS) of particles with spins \emph{moving} on a circle: the dynamical elliptic spin-Ruijsenaars system, which is also new. 
Both models are (quantum) integrable in the sense that they possess a family of conserved charges including the hamiltonians. 
The freezing procedure guarantees that the commutativity of these conserved charges is preserved when passing from the QMBS to the spin chain.

Since the $\mathit{SU}(2)$-symmetric Inozemtsev chain is a limit of our $\mathit{U}(1)$-symmetric generalisation, through our work the Inozemtsev chain, too, is embedded in the framework of freezing at last. It thus gives strong evidence for its integrability (existence of many conserved charges), although extracting explicit conserved charges from \eqref{eq:hamiltonian} requires effort, cf.\ Remark~ii in \textsection1.3.4 of \cite{lamers2022spin}. 
Moreover, our work provides a first glimpse of underlying algebraic structures via the appearance of \textit{R}-matrices. The latter depend on an extra `dynamical' parameter, not unlike suggestions of \cite{Inozemtsev:1989yq,Hal_94}. Thus, our work presents a major step towards a general theory of (quantum) integrability for long-range models with spins.

Our models differ from those of Matushko and Zotov \cite{MZ_23a,MZ_23b} in that the deformed spin interactions are built from the (face-type) dynamical elliptic \textit{R}-matrix, rather than the (vertex-type) elliptic \textit{R}-matrix of Baxter. Unlike for periodic nearest-neighbour chains, the two sets of models are not related by a face-vertex transformation. The difference has significant implications for the physical properties, even in all limits \cite{klabbers2024landscapes}.

In addition to recovering known limits, we showed that the deformed Inozemtsev chain also has two new limits. Its short-range limit is a twisted Heisenberg \textsc{xxz} chain that seems to be new and is related to the affine Temperley--Lieb algebra in the spirit of \cite{martin1994blob}, certainly warranting further investigation. Other promising directions are \textsc{rsos} specialisations, cf.~\cite{andrews1984eight}. It would also be worth investigating our novel intermediate 
generalisation of the Inozemtsev spin chain depending on an extra parameter $a'$, which sits somewhere between the latter and its deformed generalisation in Fig.~\ref{fg:spin_limits}. The fact that the parameter $a'$ disappears in all limits (including 
infinite length) makes this model rather unique, and its solution structure could shed light on the particular challenges that appear at the elliptic level.

\paragraph{Outlook.} Our work opens up many new directions.

The exact characterisation of the energies and eigenstates of our models is left for future work. The spin chain magnons, eigenstates of the (twisted) translation operator, already exhibit rich structure, making it quite non-trivial to find the dispersion relation. The eigenstates of both the isotropic Inozemtsev and deformed Haldane--Shastry chain rely on a connection to a \emph{scalar} QMBS. It is natural to investigate whether our freezing procedure can produce eigenstates for the chain from the eigenfunctions of the scalar elliptic Ruijsenaars model \cite{billey_algebraic_1998,felder_hypergeometric_2004,ruijsenaars_hilbert-schmidt_2009,langmann_construction_2022} as well, connecting it to elliptic Macdonald theory and elliptic toroidal algebras beyond $\mathfrak{gl}_1$, cf.\ \cite{konno2023elliptic}. Through suitable short-range limits, we believe this will provide a new perspective even on the well-known Bethe-ansatz solution of the isotropic Heisenberg chain.

The anisotropy of our deformed Inozemtsev chain can be set to points of special interest for condensed-matter theory, where it will simplify to yield new long-range models with e.g.\ free fermions or supersymmetry on the lattice, cf.~\cite{BLST24}. 

Our work also has implications for high-energy theory: long-range spin chains naturally appear in AdS/CFT integrability~(see \cite{Serban:2004jf,rej2012review,de2022lifting,gombor2022wrapping} and references therein), and our QMBS is closely related to supersymmetric gauge theories in five dimensions, cf.~\cite{koroteev2020quantum,gorsky2022double, konno2023elliptic}. Finally, it provides a test for the conjectured spin-version of the (quantum) `DELL' (double elliptic) system~\cite{koroteev2020quantum,gorsky2022double}.

\section*{Acknowledgements}

First and foremost we thank G.~Arutyunov for introducing us to the Inozemtsev chain and supporting us in the work that eventually led to this work, cf.~\cite{lamers2016elliptic}. We thank O.~Chalykh, G.~Felder, F.~Göhmann, P.~Koroteev, M.~Ren, H.~Rosengren, D.~Serban and M.~Volk for interest and discussions, H.~Konno for correspondence, and J.-S.~Caux, D.~Serban, A.~Sfondrini and especially B.~Doyon for feedback on drafts. We thank the organisers of the conference \textit{Integrability, Dualities and Deformations} at Humboldt-Universität zu Berlin (2022), where a key step of this work was made. JL presented this work at \textit{Integrability in Gauge and String Theory}, ETH Zürich (2023).

\paragraph{Funding information.}
The work of JL was funded by Labex Mathématique Hadamard (LMH), and in the final stage by ERC-2021-CoG\,--\,BrokenSymmetries 101044226.

\begin{appendix}
\numberwithin{equation}{section}

\section{Elliptic functions}
\label{app:ell}

Here we summarise the definitions of the elliptic functions that we need. See~\cite{klabbers2022coordinate} (where the functions $\theta$ and $\rho$ defined below were decorated with a subscript~`2') and \cite{KL_extended} for more details or the standard references \cite{DLMF,abramowitz1948handbook,whittaker1904course}.

We use the (odd) Jacobi theta function with nome $p = \E^{-N\kappa}$, which is a periodisation of a hyperbolic sine:
\begin{equation} \label{eq:theta_app}
	\theta(x) = \frac{\sinh(\kappa \, x)}{\kappa} \prod_{n=1}^{\infty}\! \frac{\sinh [\kappa \, (N \, n + x)] \sinh[\kappa \, (N \, n - x)]}{\sinh^2 (N \kappa \, n)} = \frac{\sinh(\kappa \, x)}{\kappa} + O(p^2) \, .
\end{equation}
It is the unique odd entire function with double quasiperiodicity 
\begin{equation}
\theta(x + \I\pi/\kappa)  = -\theta(x)\, \qquad \theta(x + N) = - \E^{\kappa (2x+N)} \, \theta(x)
\end{equation}
and normalisation $\theta'(0)=1$.
In terms of the Weierstra{\ss} sigma function with quasiperiods $N$ and $\I\pi/\kappa$ it reads 
\begin{equation}
	\theta(x) = \E^{\I\mspace{2mu}\kappa\,\eta_2 \mspace{2mu} x^2\!/2\pi} \, \sigma(x) \, , \qquad 
	\eta_2 = 2\,\zeta(\I\pi/2\kappa) \, .
\end{equation}
It obeys the addition formula
\begin{equation} \label{eq:addition}
\begin{aligned}
	\theta(x+y) \, \theta(x-y) \, \theta(z+w) \, \theta(z-w) = {} &\theta(x+z) \, \theta(x-z) \, \theta(y+w) \, \theta(y-w) \\ & + \theta(x+w) \, \theta(x-w) \, \theta(y+z) \, \theta(y-z) \, .
	\end{aligned}
\end{equation}
When $\kappa\to0$ we have $\theta(x) \to N \sin(\pi\,x/N)/\pi$ by the Jacobi imaginary transformation.

The prepotential is the logarithmic derivative
\begin{equation}
	\rho(x) = \frac{\theta'(x)}{\theta(x)} = \zeta(x) + \frac{\I\,\kappa\,\eta_2}{\pi} \, x = \kappa \coth(\kappa\,x) + O(p^2) \, ,
\end{equation}
with $\zeta(x) = \sigma'(x)/\sigma(x)$ the Weierstra{\ss} zeta function. It is odd and obeys $\rho(x+\I\pi/\kappa) = \rho(x)$, $\rho(x+N) = \rho(x) + 2\kappa $.

Finally, the potential is defined as the symmetric difference quotient
\begin{equation} \label{eq:pot_app}
\begin{aligned}
	V\mspace{-1mu}(x) &= -\frac{\rho(x+\eta) - \rho(x-\eta)}{\theta(2\eta)} =  \frac{A}{\mathrm{sn}[B\,(x+\eta),k]\,\mathrm{sn}[B\,(x-\eta),k]} + C \, , \\ 
	k &= \frac{\sqrt{\wp(\I\pi/2\kappa) - \wp[(N+\I\pi/\kappa)/2]}}{\sqrt{\wp(N/2) - \wp[(N+\I\pi/\kappa)/2]}} \, , 
	\end{aligned} 
\end{equation}
where the equality with Jacobi's elliptic sine $\mathrm{sn}(x,k)$, with elliptic modulus~$k$, involves constants $A,C$ (determined by	 the values at $x=0,N/2$) and $B=\sqrt{\wp(N/2) - \wp(N/2+\I\pi/2\kappa)}$. The potential is even and doubly periodic, $V(x+\I\pi/\kappa) = V(x+N) = V(x)$. The sign in \eqref{eq:pot_app} is chosen such that $V(x) \to -\rho'(x) = \wp(x) - \I \mspace{1mu}\kappa \mspace{1mu} \eta_2/\pi$ becomes the Weierstra{\ss} elliptic function as $\eta\to 0$.

\section{Deformed permutations}
\label{app:def_perm}
 
One way to obtain the dynamical \textit{R}-matrix \eqref{eq:Rdyn} is from Baxter's \textit{R}-matrix of the eight-vertex model using the face-vertex transformation~\eqref{eq:FV} \cite{baxter1973eight, TF79, FELDER1996485}. As the name of the transformation suggests, one often thinks of $\check{R}(x,a)$ as defining a `(interaction-round-the-)face' (or `\textsc{irf}') model. One can equivalently view this model as a `height model', in which case it is often called the (`elliptic' or `eight-vertex') `solid-on-solid' (or `\textsc{sos}') model, which can be described as a version of the six-vertex model where each face is decorated by a `height'. 

One face of the lattice is given a `reference' height~$a$, which determines the heights of all other faces by the spin configuration on the lines of the vertex model through the rule 
\begin{equation} \label{eq:identity}
	\tikz[baseline={([yshift=-.5*11pt*0.13-3pt]current bounding box.center)},scale=0.3,font=\footnotesize]{
		\draw[thick,gray!25] (-1,1) -- (1,1);
		\draw[->] (0,0) node[below]{$s$} -- (0,2) node[above]{$s$};
		\node at (-1,1) {$a$};
		\node at (1,1) {$b$};
	} \, , \qquad b = a - s \, ,
\end{equation}
where the line carries a spin $s = \pm1$, and $\ket{+1} \equiv \ket{\uparrow}$ and $\ket{-1} \equiv \ket{\downarrow}$. The  matrix entries of the identity correspond to
\begin{equation} \label{eq:identity_entry}
	\delta_{s,t} = \braket{t\mspace{1mu}}{\mspace{1mu}s} = 
	\tikz[baseline={([yshift=-.5*11pt*0.13-3pt]current bounding box.center)},scale=0.3,font=\footnotesize]{
		\draw[thick,gray!25] (-1,1) -- (1,1);
		\draw[->] (0,0) node[below]{$s$} -- (0,2) node[above]{$t$};
		\node at (-1,1) {$a$};
		\node at (1,1) {$b\vphantom{p}$};
	} \, , \qquad b = a -s = a-t \, ,
\end{equation}
Furthermore giving each line a spectral parameter, the generalised vertex model has vertices
\begin{equation} \label{eq:Rdyn_entry}
	\bra{t',t''} \, \check{R}(x'-x'',a) \, \ket{s',s''} = \
	\tikz[baseline={([yshift=-.5*11pt*0.4]current bounding box.center)},xscale=.6,yscale=0.3,font=\footnotesize]{
		\draw[thick,gray!25] (-.3,1.5) -- (.5,2.5) -- (1.3,1.5) -- (.5,.5) -- cycle;
		\node at (-.3,1.5) {$a$};
		\node at (.5,2.5) {$b$};
		\node at (1.3,1.5) {$c$};
		\node at (.5,.5) {$d$};
		\draw[rounded corners=2pt,->] (1,0) node[below]{$x''\mathrlap{\!,s''}$} -- (1,1) -- (0,2) -- (0,3) node[above]{$\mathllap{x''\!,\,}t'$};
		\draw[rounded corners=2pt,->] (0,0) node[below]{$\mathllap{x'\!,\,}s'$} -- (0,1) -- (1,2) -- (1,3) node[above]{$x\mathrlap{'\!,t''}$};
	} \ \ , \qquad 
	\begin{aligned} 
		b & = a - t' \, , \\
		d & = a - s' \, ,
	\end{aligned} \quad 
	c = b - t'' = d - s'' \, ,
\end{equation}
with (statistical-mechanical) weight equal to the corresponding entry of~\eqref{eq:Rdyn}. The equality on the right uses the ice rule (spin-$z$ conservation) $s'+s'' = t'+t''$ of the dynamical \textit{R}-matrix. By passing to the dual lattice, where the heights are instead attached to the vertices, one arrives at the standard \textsc{irf} picture shown in gray in \eqref{eq:identity}--\eqref{eq:Rdyn_entry}, with weight $W\Bigl(a \! \begin{array}{c}
	b \\[-.5ex] d
\end{array} c \,\Big|\, x'-x'' \Bigr)$.
One of the benefits of the generalised-vertex perspective is that the \textit{R}-matrix with entries~\eqref{eq:Rdyn_entry} is just a $4\times4$ matrix (in the spin, rather than height, basis) as in \eqref{eq:Rdyn}, i.e.\
\begin{equation} \label{eq:Rdyn_SM}
	\check{R}(x,a) = 
	\begin{pmatrix}
		\,1 & \color{gray!80}{0} & \color{gray!80}{0} & \color{gray!80}{0}\, \\
		\,\color{gray!80}{0} & f(\eta,x,\eta\,a) & f(x,\eta,\eta\,a) & \color{gray!80}{0}\, \\
		\,\color{gray!80}{0} & f(x,\eta,-\eta\,a) & f(\eta,x,-\eta\,a) & \color{gray!80}{0}\, \\
		\,\color{gray!80}{0} & \color{gray!80}{0} & \color{gray!80}{0} & 1\,
	\end{pmatrix} , 
	\quad
	f(x,y,z) = \frac{\theta(x)\,\theta(y+z)}{\theta(x+y)\,\theta(z)} \, .
\end{equation}
The price to pay is an additional parameter, $a$, that has to be shifted in the appropriate way, determined by \eqref{eq:identity_entry}. The dynamical $R$-matrix obeys the unitarity relation $\check{R}(x,a)\,\check{R}(-x,a) = 1$ and initial condition $\check{R}(0,a) = 1$. In components, unitarity reads
\begin{gather}
	\mbox{} \nonumber \\[-.65\baselineskip]
	\bra{t',t''} \, \check{R}(x''-x',a) \, \check{R}(x'-x'',a) \, \ket{s',s''} = \,
	\smash{ \tikz[baseline={([yshift=-.5*11pt*0.4]current bounding box.center)},xscale=.6,yscale=0.3,font=\footnotesize]{
			\draw[thick,gray!25] (-.3,1.5) -- (1.3,3.5) -- (.5,4.5) -- (-.3,3.5) -- (1.3,1.5) -- (.5,.5) -- cycle;
			\draw[thick,gray!35,dashed] (-.3,1.5) -- (-.3,3.5);
			\draw[thick,gray!35,dashed] (1.3,3.5) -- (1.3,1.5);
			\node at (-.3,1.5) {$a$}; \node at (-.3,3.5) {$a$};
			\node at (.5,4.5) {$b$};
			\node at (1.3,1.5) {$c$}; \node at (1.3,3.5) {$c$};
			\node at (.5,.5) {$d$};
			\node at (.5,2.5) {$e$};
			\draw[rounded corners=2pt,->] (1,0) node[below]{$x''\mathrlap{\!,s''}$} -- (1,1) -- (0,2) -- (0,3) -- (1,4) -- (1,5) node[above]{$x''\mathrlap{\!,t''}$};
			\draw[rounded corners=2pt,->] (0,0) node[below]{$\mathllap{x'\!,\,}s'$} -- (0,1) -- (1,2) -- (1,3) -- (0,4) -- (0,5) node[above]{$\mathllap{x'\!,\,}t'$};
	} } \ = \delta_{b,d} \times
	\tikz[baseline={([yshift=-.5*11pt*0.4]current bounding box.center)},xscale=.6,yscale=0.3,font=\footnotesize]{
		\draw[thick,gray!25] (-.5,1.5)--(1.5,1.5);
		\node at (-.5,1.5) {$a$}; 
		\node at (.5,1.5) {$\smash{b}\vphantom{a}$};
		\node at (1.5,1.5) {$c$};
		\draw[rounded corners=2pt,->] (1,0) node[below]{$x''\mathrlap{\!,s''}$} -- (1,3) node[above]{$x''\mathrlap{\!,t''}$};
		\draw[rounded corners=2pt,->] (0,0) node[below]{$\mathllap{x'\!,\,}s'$} -- (0,3) node[above]{$\mathllap{x'\!,\,}t'$};
	} \, = \delta_{s'\!,\,t'} \, \delta_{s''\!,\,t''} \, ,
	 \, , \nonumber \\[-.65\baselineskip]
	\mbox{} \label{eq:unitarity_entries}
\end{gather}
with $b = a - s'$ and $c = b - s''$, and where in the first diagram dashed lines join heights that are to be identified, and a sum over the spins on the two internal edges (equivalently, over the heights $e$ on the internal face) is understood. In addition, \eqref{eq:Rdyn_SM} obeys the (braid-like form of the) dynamical Yang--Baxter equation (or Gervais--Neveu--Felder equation)
\begin{equation} \label{eq:DYBEapp}
\begin{aligned} 
	\check{R}_{12}(x'-x'',a) \, & \check{R}_{23}(x-x'',a-\sigma^z_1) \, \check{R}_{12}(x-x',a) \\
	= {} & \check{R}_{23}(x-x',a-\sigma^z_1) \, \check{R}_{12}(x-x'',a) \, \check{R}_{23}(x-x',a-\sigma^z_1) \, .
	\end{aligned} 
\end{equation}
In components it reads
\begin{equation}
	\begin{aligned}
	& \bra{t,t',t''} \, \check{R}_{12}(x'-x'',a) \, \check{R}_{23}(x-x'',\,\overbrace{\!a-\sigma^z_1}^{\displaystyle \hphantom{g} = g}\,) \, \check{R}_{12}(x-x',a) \, \ket{s,s',s''} \\
	& = \
	\tikz[baseline={([yshift=-.5*11pt*0.4]current bounding box.center)},xscale=.7,yscale=0.4,font=\footnotesize]{
		\draw[thick,gray!25] (1.5,3.5) -- (-.5,1.5) -- (.5,.5) -- (2.5,2.5) -- (.5,4.5) -- (-.5,3.5) -- (1.5,1.5);
		\draw[thick,gray!35,densely dashed] (-.5,1.5) -- (-.5,3.5);
		\draw[thick,gray!25] (1.5,3.5) -- (2.5,4.5);
		\draw[thick,gray!25] (1.5,1.5) -- (2.5,.5);
		\draw[thick,gray!35,densely dashed] (2.5,4.5) -- (2.5,.5);
		\node at (-.5,1.5) {$a$};
		\node at (-.5,3.5) {$a$};
		\node at (.5,4.5) {$b$};
		\node at (1.5,3.5) {$c$}; 
		\foreach \y in {0.5,2.5,4.5} \node at (2.5,\y) {$d$};
		\node at (1.5,1.5) {$e$};
		\node at (.5,.5) {$f$};
		\node at (.5,2.5) {$g$};
		\draw[rounded corners=2pt,->] (0,0) node[below]{$\mathllap{x,\,}s\vphantom{'}$} -- (0,1) -- (2,3) -- (2,5) node[above]{$x\mathrlap{,t''}$};
		\draw[rounded corners=2pt,->] (1,0) node[below]{$x'\!,s'$} -- (1,1) -- (0,2) -- (0,3) -- (1,4) -- (1,5) node[above]{$x'\!,t'$};
		\draw[rounded corners=2pt,->] (2,0) node[below]{$x\mathrlap{''\!,s''}$} -- (2,2) -- (0,4) -- (0,5) node[above]{$\mathllap{x''\!,\,}t$};
	} \ = \
	\tikz[baseline={([yshift=-.5*11pt*0.4]current bounding box.center)},xscale=.7,yscale=0.4*.25*5,font=\footnotesize]{
		\draw[thick,gray!25] (-.3,2) -- (.5,3) -- (1.5,3) -- (2.3,2) -- (1.5,1) -- (.5,1) -- (-.3,2) -- (.7,2) (1.5,1) -- (.7,2) -- (1.5,3);
		\node at (-.3+.1,2) {$a$};
		\node at (.5,3) {$b$};
		\node at (1.5,3) {$c$};
		\node at (2.3-.1,2) {$d$};
		\node at (1.5,1) {$e$};
		\node at (.5,1-.1) {$f$};
		\node at (.7+.15,2-.05) {$g$};
		\draw[rounded corners=2pt,->] (0,0) node[below]{$\mathllap{x,\,}s\vphantom{'}$} -- (0,1.5) -- (1+.15,1.5) -- (2,2.5) -- (2,4) node[above]{$x\mathrlap{,\,t''}$};
		\draw[rounded corners=2pt,->] (1,0) node[below]{$x'\!,s'$} -- (1,1) -- (.2,2) -- (1,3) -- (1,4) node[above]{$x'\!,t'$};
		\draw[rounded corners=2pt,->] (2,0) node[below]{$x\mathrlap{''\!,s''}$} -- (2,1.5) -- (1+.15,2.5) -- (0,2.5) -- (0,4) node[above]{$\mathllap{x''\!,}t$};
	} \ \ = \ \
	\tikz[baseline={([yshift=-.5*11pt*0.4]current bounding box.center)},xscale=-.7,yscale=0.4*.25*5,font=\footnotesize]{
		\draw[thick,gray!25] (-.3,2) -- (.5,3) -- (1.5,3) -- (2.3,2) -- (1.5,1) -- (.5,1) -- (-.3,2) -- (.7,2) (1.5,1) -- (.7,2) -- (1.5,3);
		\node at (-.3+.1,2) {$d$};
		\node at (.5,3) {$c$};
		\node at (1.5,3) {$b$};
		\node at (2.3-.1,2) {$a$};
		\node at (1.5,1) {$f$};
		\node at (.5,1-.1) {$e$};
		\node at (.7+.15,2) {$h$};
		\draw[rounded corners=2pt,->] (0,0) node[below]{$x\mathrlap{''\!,s''}$} -- (0,1.5) -- (1+.15,1.5) -- (2,2.5) -- (2,4) node[above]{$\mathllap{x''\!,\,}t$};
		\draw[rounded corners=2pt,->] (1,0) node[below]{$x'\!,s'$} -- (1,1) -- (.2,2) -- (1,3) -- (1,4) node[above]{$x'\!,t'$};
		\draw[rounded corners=2pt,->] (2,0) node[below]{$\mathllap{x,\,}s\vphantom{'}$} -- (2,1.5) -- (1+.15,2.5) -- (0,2.5) -- (0,4) node[above]{$x\mathrlap{,t''}$};
	} \ = \
	\tikz[baseline={([yshift=-.5*11pt*0.4]current bounding box.center)},xscale=-.7,yscale=0.4,font=\footnotesize]{
		\draw[thick,gray!25] (1.5,3.5) -- (-.5,1.5) -- (.5,.5) -- (2.5,2.5) -- (.5,4.5) -- (-.5,3.5) -- (1.5,1.5);
		\draw[thick,gray!35,densely dashed] (-.5,1.5) -- (-.5,3.5);
		\draw[thick,gray!25] (1.5,3.5) -- (2.5,4.5);
		\draw[thick,gray!25] (1.5,1.5) -- (2.5,.5);
		\draw[thick,gray!35,densely dashed] (2.5,4.5) -- (2.5,.5);
		\node at (-.5,1.5) {$d$};
		\node at (-.5,3.5) {$d$};
		\node at (.5,4.5) {$c$};
		\node at (1.5,3.5) {$b$};
		\foreach \y in {0.5,2.5,4.5} \node at (2.5,\y) {$a$};
		\node at (1.5,1.5) {$f$};
		\node at (.5,.5) {$e$};
		\node at (.5,2.5) {$h$};
		\draw[rounded corners=2pt,->] (0,0) node[below]{$x\mathrlap{''\!,s''}$} -- (0,1) -- (2,3) -- (2,5) node[above]{$\mathllap{x''\!,\,}t$};
		\draw[rounded corners=2pt,->] (1,0) node[below]{$x'\!,s'$} -- (1,1) -- (0,2) -- (0,3) -- (1,4) -- (1,5) node[above]{$x'\!,t'$};
		\draw[rounded corners=2pt,->] (2,0) node[below]{$\mathllap{x,\,}s\vphantom{'}$} -- (2,2) -- (0,4) -- (0,5) node[above]{$x\mathrlap{,t''}$};
	} \\ 
	& = \bra{t,t',t''} \, \check{R}_{23}(x-x',\,\underbrace{\!a-\sigma^z_1}_{\displaystyle \hphantom{b} = b}\,) \, \check{R}_{12}(x-x'',a) \, \check{R}_{23}(x-x',\,\underbrace{\!a-\sigma^z_1}_{\displaystyle \hphantom{f} = f}\,) \, \ket{s,s',s''} \, ,
	\end{aligned}
\end{equation}
where sums over spins on the three internal lines (equivalently, over the height $g$ or $h$ of the internal face) are again understood. The resulting algebraic structure is Felder's elliptic quantum group~\cite{felder1994elliptic}.

Now consider a row of $N$ vertical lines in the generalised vertex model. The deformed permutation \eqref{eq:deformed_permutation} similarly encodes the vertex
\begin{equation} \label{eq:deformed_permutation_entries}
	\bra{t_1,\dots,t_N} \, P_{i,i+1}(x' - x'') \, \ket{s_1,\dots,s_N} = 
	\tikz[baseline={([yshift=-.5*11pt*0.3]current bounding box.center)},xscale=.5,yscale=.4,font=\footnotesize]{
		\draw[->] (0,0) node[below] {$s_1$} -- (0,3) node[above] {$t_1$};
		\foreach \x in {-1,...,1} \draw (.75+.2*\x,1.5) node{$\cdot\mathstrut$};	
		\draw[->] (1.5,0) -- (1.5,3);
		\draw[rounded corners=2pt,->] (4,0) node[below, yshift=.05cm]{$x''\mathrlap{\!, s_{i+1}}$} -- (4,1) -- (3,2) -- (3,3) node[above, yshift=-.03cm]{$\mathllap{x''\!,\,} t_i$};
		\draw[rounded corners=2pt,->] (3,0) node[below, yshift=.05cm]{$\mathllap{x'\!,\,} s_i$} -- (3,1) -- (4,2) -- (4,3) node[above, yshift=-.03cm]{$\!\!x'\mathrlap{\!,t_{i+1}}$};
		\draw[->] (5.5,0) -- (5.5,3);
		\foreach \x in {-1,...,1} \draw (6.25+.2*\x,1.5) node{$\cdot\mathstrut$};	
		\draw[->] (7,0) node[below] {$s_N$} -- (7,3) node[above] {$t_N$};
		\node at (-.5,1.5) {$a$};
		\node at (2.45,1.3) {$a_{i-1}$};
		\node at (3.5,2.5) {$a''_i$};
		\node at (4.65,1.3) {$a_{i+1}$};
		\node at (3.5,.5) {$a'_i$};
		\node at (7.5,1.5) {$a_N$};
	} \, ,
\end{equation}
where we omitted the spectral parameters attached to all non-crossing lines to avoid cluttering, and the heights are
\begin{equation}
	a_0 = a \, , \qquad 
	a_j = a_{j-1} - s_j \quad (j \neq i,i+1) \, , \qquad
	\begin{aligned} 
		a''_i & = a_{i-1} - t_i \, , \\
		a'_i & = a_{i-1} - s_i \, ,
	\end{aligned} \qquad 
	a_{i+1} = a''_i - t_{i+1} = a'_i - s_{i+1} \, .
\end{equation}
The vertex \eqref{eq:deformed_permutation_entries} corresponds to a single matrix entry of $P_{i,i+1}(x)$. The whole matrix can be written as in \eqref{eq:deformed_permutation}, i.e. 
\begin{equation} \label{eq:deformed_permutation_SM}
	P_{i,i+1}(x)  = \check{R}_{i,i+1}\bigl(x,a-(\sigma^z_1+\dots+\sigma^z_{i-1})\bigr) 
\end{equation}
On the usual spin (`computational') basis this notation means
\begin{equation} \label{eq:meaning_shifts}
	\begin{aligned}
	\! P_{i,i+1}(x) \, \ket{s_1,\dots,s_N} = {} & \ket{s_1,\dots,s_{i-1}} \otimes \Bigl( \check{R}\bigl(x,a-\textstyle\sum_{k=1}^{i-1} s_k\bigr)\, \ket{s_i,s_{i+1}} \Bigr) \otimes \ket{s_{i+1},\dots,s_N} \, .	
	\end{aligned}
\end{equation}
We stress once more that the dynamical parameter of the \textit{R}-matrix in \eqref{eq:deformed_permutation_SM}--\eqref{eq:meaning_shifts} is shifted by (twice) the spin-$z$ to the left of the 
	$\tikz[baseline={([yshift=-2*11pt*.15]current bounding box.center)},xscale=.3,yscale=.15]{
		\draw[rounded corners=2pt,->] (1,-.1) -- (1,.5) -- (0,1.5) -- (0,2.3);
		\draw[rounded corners=2pt,->] (0,-.1) -- (0,.5) -- (1,1.5) -- (1,2.3);
	}\,$
in agreement with \eqref{eq:deformed_permutation_entries}.
Projecting on $\bra{t_1,\dots,t_N}$ we recover \eqref{eq:deformed_permutation_entries}. 

Thanks to \eqref{eq:DYBEapp}, the deformed permutations obey the (braid-like) Yang--Baxter equation
\begin{equation}
	\begin{aligned} 
		P_{i,i+1}(x-y) \, P_{i+1\mspace{-1mu},\mspace{1mu}i+2}(x) \, P_{i,i+1}(y) 
		= {}  P_{i+1\mspace{-1mu},\mspace{1mu}i+2}(y) \, P_{i,i+1}(x) \, P_{i+1\mspace{-1mu},\mspace{1mu}i+2}(x-y) \, ,
	\end{aligned}
\end{equation}
as well as the commutativity $[P_{i,i+1}(x),P_{j\mspace{-1mu},\mspace{1mu}j+1}(y)]=0$ for $|i-j| > 1$. They moreover inherit the unitarity relation
\begin{equation} \label{eq:unitarity}
	P_{i,i+1}(-x)\, P_{i,i+1}(x) = 1 \, .
\end{equation}
with `initial condition' $P_{i,i+1}(0) = 1$. According to \eqref{eq:unitarity}, swapping twice is the identity. That is, taking into account that the parameters follow the lines, the deformed permutations square (appropriately interpreted) to the identity. This can be made precise by introducing the coordinate permutation $s_{ij} : x_i \leftrightarrow x_j$. Consider the deformed total permutation
\begin{equation} \label{eq:total_perm}
	P_{i,i+1}^\text{tot} = s_{i,i+1} \, P_{i,i+1}(x_i - x_{i+1}) \, .
\end{equation}
It permutes particles, i.e.\ spins \emph{and} coordinates. (Since parameters should follow lines in diagrams, one could draw it as 
$\tikz[baseline={([yshift=-2*11pt*0.15]current bounding box.center)},xscale=0.3,yscale=0.15]{
	\draw[->] (1,0) -- (1,2.4);
	\draw[->] (2,0) -- (2,2.4);
	\draw (1,1) -- (2,1);}$\,.) 
Now \eqref{eq:DYBEapp} becomes the braid relation
\begin{equation} 
	P_{i,i+1}^\text{tot} \, P_{i+1\mspace{-1mu},\mspace{1mu}i+2}^\text{tot} \, P_{i,i+1}^\text{tot} 
	= P_{i+1\mspace{-1mu},\mspace{1mu}i+2}^\text{tot} \, P_{i,i+1}^\text{tot} \, P_{i+1\mspace{-1mu},\mspace{1mu}i+2}^\text{tot} \, ,
\end{equation}
we have $\bigl[P_{i,i+1}^\text{tot},P_{j,j+1}^\text{tot}\bigr] = 0$ for $|i-j|>1$, and \eqref{eq:unitarity} reads
\begin{equation}
	\bigl(P_{i,i+1}^\text{tot}\bigr)^{\!2} = 1 \, .
\end{equation}
These are the relations of the permutation group. In the isotropic limit $\eta \to 0$ we recover the standard particle permutation, $P_{i,i+1}^\text{tot} \to s_{i,i+1} \, P_{i,i+1}$. For general $\eta$, \eqref{eq:total_perm} depends on all parameters.

\section{Deformed nearest-neighbour exchange}
\label{app:def_nn_exchange} 

The deformed spin exchange
\begin{equation} \label{eq:pair_interaction_app}
	E(x,a) = \frac{1}{\theta(\eta) \, V\mspace{-1mu}(x)} \, \check{R}(-x,a) \, \check{R}'(x,a) = \tikz[baseline={([yshift=-.5*11pt*0.4]current bounding box.center)},xscale=.4,yscale=0.2,font=\footnotesize]{
		\draw[->] (0,0) node[below]{$x'$} -- (0,3) node[above]{$x'$};
		\draw[->] (1,0) node[below]{$\,\,x''$} -- (1,3) node[above]{$\,\,x''$};
		\draw[style={decorate, decoration={zigzag,amplitude=.5mm,segment length=1mm}}] (0,1.5) -- (1,1.5);
		\node at (-.35,1.5) {$a$};
	} \! , \quad
	 \check{R}'(x,a) \equiv \partial_x \check{R}(x,a) \, , 
	\quad x = x' - x'' \, ,
\end{equation}
is nothing but a normalised logarithmic derivative of the dynamical \textit{R}-matrix, $\partial \log \check{R} = \check{R}^{-1} \check{R}'$, mirroring the local hamiltonians of Heisenberg chains. 
As an explicit $4\times 4$ matrix it reads
\begin{equation} 
	\theta(\eta) \, V(x) \, E(x,a) = \check{R}(-x,a) \, \check{R}'(x,a) = 
	\begin{pmatrix}
		0 & \color{gray!80}{0} & \color{gray!80}{0} & \color{gray!80}{0} \\
		\color{gray!80}{0} & \alpha(x,\eta\,a) & \beta(x,\eta\,a) & \color{gray!80}{0} \\
		\color{gray!80}{0} & \beta(x,-\eta\,a)\! & \alpha(x,-\eta\,a)\! & \color{gray!80}{0} \\
		\color{gray!80}{0} & \color{gray!80}{0} & \color{gray!80}{0} & 0
	\end{pmatrix} \, ,
\end{equation}
where the first equality uses the unitarity $\check{R}(x,a)^{-1} = \check{R}(-x,a)$, and the coefficients are
\begin{equation} \label{eq:alpha,beta}
	\begin{aligned}
	\alpha(x,a) = {} & 
	f(\eta,x,a) \, f(\eta,-x,a) \, \bigl(\rho(x+a)-\rho(x)\bigr) - \bigl(\rho(x+\eta)-\rho(x)\bigr) \\
	= {} & f(\eta,x,a) \, f(\eta,-x,a) \, \rho(x+a) + f(x,\eta,a)\,f(-x,\eta,-a) \, \rho(x) - \rho(x+\eta) \, , \\
	\beta(x,a) = {} & f(x,\eta,a)\,f(\eta,-x,a)\,\bigl(\rho(x)-\rho(x-a)\bigr) \, . 
	\end{aligned}
\end{equation}

Its entries can be interpreted like in \eqref{eq:unitarity_entries}: if 
`$\mspace{-2mu}$\raisebox{-1.5pt}{\rotatebox{30}{$\circledast$}}$\mspace{-2mu}$' 
marks the derivative of $\check{R}'$,
\begin{equation} \label{eq:pair_interaction_entry}
	\bra{t',t''} \, E(x'-x'',a) \, \ket{s',s''} = \
	\tikz[baseline={([yshift=-.5*11pt*0.4]current bounding box.center)},xscale=.7,yscale=0.4,font=\footnotesize]{
		\draw[thick,gray!25] (-.3,1.5) -- (.5,2.5) -- (1.3,1.5) -- (.5,.5) -- cycle;
		\node at (-.35,1.5) {$a$};
		\node at (.5,2.5) {$b$};
		\node at (1.35,1.5) {$c$};
		\node at (.5,.5) {$d$};
		\draw[->] (0,0) node[below]{$\mathllap{x'\!,\,}s'$} -- (0,3) node[above]{$\mathllap{x'\!,\,}t'$};
		\draw[->] (1,0) node[below]{$x''\mathrlap{\!,s''}$} -- (1,3) node[above]{$x''\mathrlap{\!,t''}$};
		\draw[style={decorate, decoration={zigzag,amplitude=.5mm,segment length=1mm}}] (0,1.5) -- (1,1.5);
	} 
	\ = \, \frac{1}{\theta(\eta) \, V\mspace{-1mu}(x'-x'')} \, 
	\tikz[baseline={([yshift=-.5*11pt*0.4]current bounding box.center)},xscale=.7,yscale=0.4,font=\footnotesize]{
		\draw[thick,gray!25] (-.3,1.5) -- (1.3,3.5) -- (.5,4.5) -- (-.3,3.5) -- (1.3,1.5) -- (.5,.5) -- cycle;
		\draw[thick,gray!35,dashed] (-.3,1.5) -- (-.3,3.5);
		\draw[thick,gray!35,dashed] (1.3,3.5) -- (1.3,1.5);
		\node at (.5,1.5cm-.5pt) {\rotatebox{30}{$\circledast$}};
		\node at (-.3,1.5) {$a$}; \node at (-.3,3.5) {$a$};
		\node at (.5,4.5) {$b$};
		\node at (1.3,1.5) {$c$}; \node at (1.3,3.5) {$c$};
		\node at (.5,.5) {$d$};
		\node at (.5,2.5) {$e$};
		\draw[rounded corners=2pt,->] (1,0) node[below]{$x''\mathrlap{\!,s''}$} -- (1,1) -- (0,2) -- (0,3) -- (1,4) -- (1,5) node[above]{$x''\mathrlap{\!,t''}$};
		\draw[rounded corners=2pt,->] (0,0) node[below]{$\mathllap{x'\!,\,}s'$} -- (0,1) -- (1,2) -- (1,3) -- (0,4) -- (0,5) node[above]{$\mathllap{x'\!,\,}t'$};
	}
	\ , \quad 
	\begin{aligned} 
		b & = a - t' \, , \\
		d & = a - s' \, ,
	\end{aligned}
\end{equation}
with $ c = b - t'' = d - s''$. 

To evaluate the isotropic limit of $E(x,a)$ note that for $\eta\to 0$
\begin{equation}
	f(\eta, x, \eta\,a) = \frac{\theta(\eta) \,\theta(x+\eta\,a)}{\theta(\eta + x) \, \theta(\eta\,a)} \to \frac{1}{a} \, , \quad
	f(x,\eta, \eta\,a) = \frac{\theta(x)\,\theta(\eta + \eta\,a)}{\theta(x+\eta) \, \theta(\eta\,a)} \to \frac{a+1}{a} \, .
\end{equation}
Hence
\begin{equation}
	\begin{gathered}
	\frac{\alpha(x,\eta\,a)}{\theta(\eta)} \to
	\frac{1}{a^2} \, a \, \rho'(x) - \rho'(x) = \biggl( 1 - \frac{1}{a}\biggr) \, \bar{V}^\text{Ino}(x) \, , \\
	\frac{\beta(x,\eta\,a)}{\theta(\eta)} \to \frac{a+1}{a^2} \, a \, \rho'(x) = -\biggl( 1 + \frac{1}{a} \biggr) \, \bar{V}^\text{Ino}(x) \, ,
	\end{gathered}
	\qquad \eta\to 0 \, ,
\end{equation}
where we recall $\bar{V}^\text{Ino}(x) = -\rho'(x)$. Letting $a\to -\I \, \infty$ we see that $E(x,a)$ reduces to $1-P$.

\end{appendix}

\bibliography{bibliography.bib}

\begin{thebibliography}{10}
\providecommand{\url}[1]{\texttt{#1}}
\providecommand{\urlprefix}{URL }
\expandafter\ifx\csname urlstyle\endcsname\relax
  \providecommand{\doi}[1]{doi:\discretionary{}{}{}#1}\else
  \providecommand{\doi}{doi:\discretionary{}{}{}\begingroup
  \urlstyle{rm}\Url}\fi
\providecommand{\eprint}[2][]{\url{#2}}

\bibitem{jurcevic2014quasiparticle}
P.~Jurcevic, B.~P. Lanyon, P.~Hauke, C.~Hempel, P.~Zoller, R.~Blatt and C.~F.
  Roos,
\newblock \emph{Quasiparticle engineering and entanglement propagation in a
  quantum many-body system},
\newblock Nature \textbf{511}(7508), 202 (2014),
\newblock \doi{10.1038/nature13461},
\newblock \eprint{1401.5387}.

\bibitem{zhang2017observation}
J.~Zhang, P.~W. Hess, A.~Kyprianidis, P.~Becker, A.~Lee, J.~Smith, G.~Pagano,
  I.-D. Potirniche, A.~C. Potter, A.~Vishwanath \emph{et~al.},
\newblock \emph{Observation of a discrete time crystal},
\newblock Nature \textbf{543}(7644), 217 (2017),
\newblock \doi{10.1038/nature21413},
\newblock \eprint{1609.08684}.

\bibitem{RevModPhys.93.025001}
C.~Monroe, W.~C. Campbell, L.-M. Duan, Z.-X. Gong, A.~V. Gorshkov, P.~W. Hess,
  R.~Islam, K.~Kim, N.~M. Linke, G.~Pagano, P.~Richerme, C.~Senko
  \emph{et~al.},
\newblock \emph{Programmable quantum simulations of spin systems with trapped
  ions},
\newblock Rev. Mod. Phys. \textbf{93}, 025001 (2021),
\newblock \doi{10.1103/RevModPhys.93.025001},
\newblock \eprint{1912.07845}.

\bibitem{zeiher2017coherent}
J.~Zeiher, J.-Y. Choi, A.~Rubio-Abadal, T.~Pohl, R.~Van~Bijnen, I.~Bloch and
  C.~Gross,
\newblock \emph{Coherent many-body spin dynamics in a long-range interacting
  {I}sing chain},
\newblock Phys. Rev. X \textbf{7}(4), 041063 (2017),
\newblock \doi{10.1103/PhysRevX.7.041063},
\newblock \eprint{1705.08372}.

\bibitem{PhysRevX.11.031016}
M.~C. Tran, A.~Y. Guo, A.~Deshpande, A.~Lucas and A.~V. Gorshkov,
\newblock \emph{Optimal state transfer and entanglement generation in power-law
  interacting systems},
\newblock Phys. Rev. X \textbf{11}, 031016 (2021),
\newblock \doi{10.1103/PhysRevX.11.031016},
\newblock \eprint{2010.02930}.

\bibitem{linke2017experimental}
N.~M. Linke, D.~Maslov, M.~Roetteler, S.~Debnath, C.~Figgatt, K.~A. Landsman,
  K.~Wright and C.~Monroe,
\newblock \emph{Experimental comparison of two quantum computing
  architectures},
\newblock PNAS \textbf{114}(13), 3305 (2017),
\newblock \doi{10.1073/pnas.1618020114},
\newblock \eprint{1702.01852}.

\bibitem{10.1063/1.5088164}
C.~D. Bruzewicz, J.~Chiaverini, R.~McConnell and J.~M. Sage,
\newblock \emph{{Trapped-ion quantum computing: Progress and challenges}},
\newblock Appl. Phys. Rev. \textbf{6}(2), 021314 (2019),
\newblock \doi{10.1063/1.5088164},
\newblock \eprint{1904.04178}.

\bibitem{defenu2023long}
N.~Defenu, T.~Donner, T.~Macr{\`\i}, G.~Pagano, S.~Ruffo and A.~Trombettoni,
\newblock \emph{Long-range interacting quantum systems},
\newblock Rev. Mod. Phys. \textbf{95}(3), 035002 (2023),
\newblock \doi{10.1103/RevModPhys.95.035002},
\newblock \eprint{2109.01063}.

\bibitem{PhysRevLett.109.025303}
D.~Peter, S.~M\"uller, S.~Wessel and H.~P. B\"uchler,
\newblock \emph{Anomalous behavior of spin systems with dipolar interactions},
\newblock Phys. Rev. Lett. \textbf{109}, 025303 (2012),
\newblock \doi{10.1103/PhysRevLett.109.025303},
\newblock \eprint{1203.1624}.

\bibitem{PhysRevLett.111.207202}
P.~Hauke and L.~Tagliacozzo,
\newblock \emph{Spread of correlations in long-range interacting quantum
  systems},
\newblock Phys. Rev. Lett. \textbf{111}, 207202 (2013),
\newblock \doi{10.1103/PhysRevLett.111.207202},
\newblock \eprint{1304.7725}.

\bibitem{gong2016kaleidoscope}
Z.-X. Gong, M.~F. Maghrebi, A.~Hu, M.~Foss-Feig, P.~Richerme, C.~Monroe and
  A.~V. Gorshkov,
\newblock \emph{Kaleidoscope of quantum phases in a long-range interacting
  spin-1 chain},
\newblock Phys. Rev. B \textbf{93}(20), 205115 (2016),
\newblock \doi{10.1103/PhysRevB.93.205115},
\newblock \eprint{1510.02108}.

\bibitem{haldane1988exact}
F.~Haldane,
\newblock \emph{Exact {J}astrow--{G}utzwiller resonating-valence-bond ground
  state of the spin-$1/2$ antiferromagnetic {H}eisenberg chain with $1/r^2$
  exchange},
\newblock Phys. Rev. Lett. \textbf{60}(7), 635 (1988),
\newblock \doi{https://doi.org/10.1103/PhysRevLett.60.635}.

\bibitem{shastry1988exact}
B.~Shastry,
\newblock \emph{Exact solution of an $s=1/2$ {H}eisenberg antiferromagnetic
  chain with long-ranged interactions},
\newblock Phys. Rev. Lett. \textbf{60}(7), 639 (1988),
\newblock \doi{https://doi.org/10.1103/PhysRevLett.60.639}.

\bibitem{PhysRev.187.732}
D.~J. Thouless,
\newblock \emph{Long-range order in one-dimensional {I}sing systems},
\newblock Phys. Rev. \textbf{187}, 732 (1969),
\newblock \doi{10.1103/PhysRev.187.732}.

\bibitem{grass2014trapped}
T.~Gra{\ss} and M.~Lewenstein,
\newblock \emph{Trapped-ion quantum simulation of tunable-range {H}eisenberg
  chains},
\newblock EPJ Quantum Technol. \textbf{1}(1), 1 (2014),
\newblock \doi{10.1186/epjqt8},
\newblock \eprint{1401.6414}.

\bibitem{Hal_91a}
F.~Haldane,
\newblock \emph{``{S}pinon gas'' description of the {$S=1/2$} {H}eisenberg
  chain with inverse-square exchange: Exact spectrum and thermodynamics},
\newblock Phys. Rev. Lett. \textbf{66}, 1529 (1991),
\newblock \doi{10.1103/PhysRevLett.66.1529}.

\bibitem{Hal_94}
F.~Haldane,
\newblock \emph{Physics of the ideal semion gas: spinons and quantum symmetries
  of the integrable {Haldane--Shastry} spin chain},
\newblock In A.~Okiji and N.~Kawakami, eds., \emph{Correlation effects in
  low-dimensional electron systems}, vol. 118. Springer,
\newblock \doi{10.1007/978-3-642-85129-2_1} (1994), \eprint{cond-mat/9401001}.

\bibitem{HH+_92}
F.~Haldane, Z.~Ha, J.~Talstra, D.~Bernard and V.~Pasquier,
\newblock \emph{Yangian symmetry of integrable quantum chains with long-range
  interactions and a new description of states in conformal field theory},
\newblock Phys. Rev. Lett. \textbf{69}, 2021 (1992),
\newblock \doi{10.1103/PhysRevLett.69.2021}.

\bibitem{bernard1994spinons}
D.~Bernard, V.~Pasquier and D.~Serban,
\newblock \emph{Spinons in conformal field theory},
\newblock Nucl. Phys. B \textbf{428}(3), 612 (1994),
\newblock \doi{10.1016/0550-3213(94)90366-2},
\newblock \eprint{hep-th/9404050}.

\bibitem{bouwknegt1994spinon}
P.~Bouwknegt, A.~W. Ludwig and K.~Schoutens,
\newblock \emph{Spinon bases, {Y}angian symmetry and fermionic representations
  of {V}irasoro characters in conformal field theory},
\newblock Phys. Lett. B \textbf{338}(4), 448 (1994),
\newblock \doi{10.1016/0370-2693(94)90799-4},
\newblock \eprint{hep-th/9406020}.

\bibitem{bouwknegt1996n}
P.~Bouwknegt and K.~Schoutens,
\newblock \emph{The $\widehat{SU(n)}_1$ {WZW} models: spinon decomposition and
  {Y}angian structure},
\newblock Nucl. Phys. B \textbf{482}(1-2), 345 (1996),
\newblock \doi{10.1016/S0550-3213(96)00565-2},
\newblock \eprint{hep-th/9607064}.

\bibitem{Inozemtsev:1989yq}
V.~Inozemtsev,
\newblock \emph{On the connection between the one-dimensional $s=1/2$
  {H}eisenberg chain and {Haldane--Shastry} model},
\newblock J. Stat. Phys. \textbf{59}(5-6), 1143 (1990),
\newblock \doi{10.1007/BF01334745}.

\bibitem{klabbers2022coordinate}
R.~Klabbers and J.~Lamers,
\newblock \emph{How coordinate {B}ethe ansatz works for {I}nozemtsev model},
\newblock Commun. Math. Phys. \textbf{390}(2), 827 (2022),
\newblock \doi{https://doi.org/10.1007/s00220-021-04281-x},
\newblock \eprint{2009.14513}.

\bibitem{inozemtsev1996invariants}
V.~Inozemtsev,
\newblock \emph{Invariants of linear combinations of transpositions},
\newblock Lett. Math. Phys. \textbf{36}(1), 55 (1996),
\newblock \doi{10.1007/BF00403251}.

\bibitem{dittrich2008commutativity}
J.~Dittrich and V.~Inozemtsev,
\newblock \emph{The commutativity of integrals of motion for quantum spin
  chains and elliptic functions identities},
\newblock Reg. Chaot. Dyn. \textbf{13}(1), 19 (2008),
\newblock \doi{10.1007/s11819-008-1003-3},
\newblock \eprint{0711.1973}.

\bibitem{Chalykh24u}
O.~Chalykh,
\newblock \emph{Integrability of the {I}nozemtsev spin chain}
  \eprint{2407.03276}.

\bibitem{bernard1993yang}
D.~Bernard, M.~Gaudin, F.~Haldane and V.~Pasquier,
\newblock \emph{Yang--{B}axter equation in long-range interacting systems},
\newblock J. Phys. A: Math. Gen. \textbf{26}(20), 5219 (1993),
\newblock \doi{https://doi.org/10.1088/0305-4470/26/20/010},
\newblock \eprint{hep-th/9301084}.

\bibitem{Ugl_95u}
D.~Uglov,
\newblock \emph{The trigonometric counterpart of the {H}aldane--{S}hastry
  model}  (1995),
\newblock \eprint{hep-th/9508145}.

\bibitem{Lam_18}
J.~Lamers,
\newblock \emph{Resurrecting the partially isotropic {H}aldane--{S}hastry
  model},
\newblock Phys. Rev. B. \textbf{97}, 214416 (2018),
\newblock \doi{https://doi.org/10.1103/PhysRevB.97.214416},
\newblock \eprint{1801.05728}.

\bibitem{lamers2022spin}
J.~Lamers, V.~Pasquier and D.~Serban,
\newblock \emph{Spin-{R}uijsenaars, \textit{q}-deformed {H}aldane--{S}hastry
  and {M}acdonald polynomials},
\newblock Commun. Math. Phys. \textbf{393}, 61 (2022),
\newblock \doi{10.1007/s00220-022-04318-9},
\newblock \eprint{2004.13210}.

\bibitem{lamers2022fermionic}
J.~Lamers and D.~Serban,
\newblock \emph{From fermionic spin-{Calogero--Sutherland} models to the
  {Haldane--Shastry} chain by freezing},
\newblock J. Phys. A: Math. Theor. \textbf{57}, 235205 (2024),
\newblock \eprint{2212.01373}.

\bibitem{Pol_93}
A.~Polychronakos,
\newblock \emph{Lattice integrable systems of {H}aldane--{S}hastry type},
\newblock Phys. Rev. Lett. \textbf{70}, 2329 (1993),
\newblock \doi{10.1103/physrevlett.70.2329},
\newblock \eprint{hep-th/9210109}.

\bibitem{TH_95}
J.~Talstra and F.~Haldane,
\newblock \emph{Integrals of motion of the {H}aldane--{S}hastry model},
\newblock J. Phys. A: Math. Gen. \textbf{28}, 2369 (1995),
\newblock \doi{10.1088/0305-4470/28/8/027},
\newblock \eprint{cond-mat/9411065}.

\bibitem{drinfel1986degenerate}
V.~Drinfel'd,
\newblock \emph{Degenerate affine {H}ecke algebras and {Y}angians},
\newblock Funct. Anal. Appl. \textbf{20}(1), 58 (1986),
\newblock \doi{https://doi.org/10.1007/BF01077318}.

\bibitem{Inozemtsev_1995}
V.~Inozemtsev,
\newblock \emph{On the spectrum of $s=1/2$ {XXX} {H}eisenberg chain with
  elliptic exchange},
\newblock J. Phys. A: Math. Gen. \textbf{28}(16), L439 (1995),
\newblock \doi{10.1088/0305-4470/28/16/004},
\newblock \eprint{cond-mat/9504096}.

\bibitem{felder1997elliptic}
G.~Felder and A.~Varchenko,
\newblock \emph{Elliptic quantum groups and {R}uijsenaars models},
\newblock J. Stat. Phys. \textbf{89}(5), 963 (1997),
\newblock \doi{https://doi.org/10.1007/BF02764216},
\newblock \eprint{q-alg/9704005}.

\bibitem{felder1994elliptic}
G.~Felder,
\newblock \emph{Elliptic quantum groups},
\newblock In D.~Iagolnitzer, ed., \emph{Proc.\ ICMP Paris 1994}, p. 211.
  International Press (1995), \eprint{hep-th/9412207}.

\bibitem{MZ_23b}
M.~Matushko and A.~Zotov,
\newblock \emph{Elliptic generalization of integrable $q$-deformed
  {H}aldane--{S}hastry long-range spin chain},
\newblock Nonlin. \textbf{36}, 319 (2023),
\newblock \doi{10.1088/1361-6544/aca510},
\newblock \eprint{2202.01177}.

\bibitem{klabbers2024landscapes}
R.~Klabbers and J.~Lamers,
\newblock \emph{Landscapes of integrable long-range spin chains}  (2024),
\newblock \eprint{2405.09718}.

\bibitem{KL_extended}
R.~Klabbers and J.~Lamers,
\newblock \emph{Freezing elliptic quantum many-body systems with spins},
\newblock In preparation.

\bibitem{arnaudon2000towards}
D.~Arnaudon, J.~Avan, L.~Frappat, E.~Ragoucy and M.~Rossi,
\newblock \emph{Towards a cladistics of double {Y}angians and elliptic
  algebras},
\newblock J. Phys. A: Math. Gen. \textbf{33}(36), 6279 (2000),
\newblock \eprint{math/9906189}.

\bibitem{Baxter1972193}
R.~J. Baxter,
\newblock \emph{Partition function of the eight-vertex lattice model},
\newblock Ann. Phys. \textbf{70}(1), 193 (1972),
\newblock \doi{10.1016/0003-4916(72)90335-1}.

\bibitem{baxter1973eight}
R.~Baxter,
\newblock \emph{Eight-vertex model in lattice statistics and one-dimensional
  anisotropic {H}eisenberg chain. {II}. {E}quivalence to a generalized ice-type
  lattice model},
\newblock Ann. Phys. \textbf{76}(1), 25 (1973),
\newblock \doi{10.1016/0003-4916(73)90440-5}.

\bibitem{jimbo1999quasi}
M.~Jimbo, H.~Konno, S.~Odake and J.~I. Shiraishi,
\newblock \emph{Quasi-hopf twistors for elliptic quantum groups},
\newblock Transf. Groups \textbf{4}, 303 (1999),
\newblock \doi{10.1007/BF01238562},
\newblock \eprint{q-alg/9712029}.

\bibitem{MZ_23a}
M.~Matushko and A.~Zotov,
\newblock \emph{Anisotropic spin generalization of elliptic
  {M}acdonald--{R}uijsenaars operators and {$R$}-matrix identities},
\newblock {Ann. H. Poincaré} \textbf{2023} (2023),
\newblock \doi{10.1007/s00023-023-01316-y},
\newblock \eprint{2201.05944}.

\bibitem{martin1993algebraic}
P.~Martin and H.~Saleur,
\newblock \emph{On an algebraic approach to higher dimensional statistical
  mechanics},
\newblock Commun. Math. Phys. \textbf{158}(1), 155 (1993),
\newblock \doi{https://doi.org/10.1007/BF02097236}.

\bibitem{belletete2017correspondence}
J.~Bellet{\^e}te, A.~M. Gainutdinov, J.~L. Jacobsen, H.~Saleur and R.~Vasseur,
\newblock \emph{On the correspondence between boundary and bulk lattice models
  and (logarithmic) conformal field theories},
\newblock J. Phys. A: Math. Theor. \textbf{50}(48), 484002 (2017),
\newblock \doi{10.1088/1751-8121/aa902b},
\newblock \eprint{1705.07769}.

\bibitem{pasquier1990common}
V.~Pasquier and H.~Saleur,
\newblock \emph{Common structures between finite systems and conformal field
  theories through quantum groups},
\newblock Nucl. Phys. B \textbf{330}(2-3), 523 (1990),
\newblock \doi{10.1016/0550-3213(90)90122-T}.

\bibitem{andrews1984eight}
G.~E. Andrews, R.~J. Baxter and P.~J. Forrester,
\newblock \emph{Eight-vertex {SOS} model and generalized
  {R}ogers--{R}amanujan-type identities},
\newblock J. Stat. Phys. \textbf{35}(3-4), 193 (1984),
\newblock \doi{https://doi.org/10.1007/BF01014383}.

\bibitem{martin1994blob}
P.~Martin and H.~Saleur,
\newblock \emph{The blob algebra and the periodic {T}emperley--{L}ieb algebra},
\newblock Lett. Math. Phys. \textbf{30}, 189 (1994),
\newblock \doi{10.1007/BF00805852}.

\bibitem{filali2011spin}
G.~Filali and N.~Kitanine,
\newblock \emph{Spin chains with non-diagonal boundaries and trigonometric
  {SOS} model with reflecting end},
\newblock SIGMA \textbf{7}, 012 (2011),
\newblock \doi{10.3842/SIGMA.2011.012},
\newblock \eprint{1011.0660}.

\bibitem{Che_94a}
I.~Cherednik,
\newblock \emph{Induced representations of double affine {H}ecke algebras and
  applications},
\newblock Math. Res. Lett. \textbf{1}(3), 319 (1994),
\newblock \doi{10.4310/MRL.1994.v1.n3.a4}.

\bibitem{hikami1993integrability}
K.~Hikami and M.~Wadati,
\newblock \emph{Integrability of {C}alogero--{M}oser spin system},
\newblock J. Phys. Soc. Japan \textbf{62}(2), 469 (1993),
\newblock \doi{10.1143/JPSJ.62.469}.

\bibitem{krichever1994spin}
I.~Krichever, O.~Babelon, E.~Billey and M.~Talon,
\newblock \emph{Spin generalization of the {C}alogero--{M}oser system and the
  matrix {KP} equation}  (1994),
\newblock \eprint{hep-th/9411160}.

\bibitem{Rui_87}
S.~N.~M. Ruijsenaars,
\newblock \emph{Complete integrability of relativistic {C}alogero--{M}oser
  systems and elliptic function identities},
\newblock Commun. Math. Phys. \textbf{110}, 191 (1987),
\newblock \doi{10.1007/BF01207363}.

\bibitem{ruijsenaars1986new}
S.~N.~M. Ruijsenaars and H.~Schneider,
\newblock \emph{A new class of integrable systems and its relation to
  solitons},
\newblock Ann. Phys. \textbf{170}(2), 370 (1986).

\bibitem{billey_algebraic_1998}
E.~Billey,
\newblock \emph{Algebraic nested {Bethe} ansatz for the elliptic {Ruijsenaars}
  model} (1998), \eprint{math/9806068}.

\bibitem{felder_hypergeometric_2004}
G.~Felder and A.~Varchenko,
\newblock \emph{Hypergeometric theta functions and elliptic {Macdonald}
  polynomials},
\newblock Int. Math. Res. Not. \textbf{2004}(21), 1037 (2004),
\newblock \doi{10.1155/S1073792804132893},
\newblock \eprint{math/0309452}.

\bibitem{ruijsenaars_hilbert-schmidt_2009}
S.~N.~M. Ruijsenaars,
\newblock \emph{Hilbert--{Schmidt} operators vs. integrable systems of elliptic
  {Calogero}--{Moser} type {I}. {The} eigenfunction identities},
\newblock Commun. Math. Phys. \textbf{286}(2), 629 (2009),
\newblock \doi{10.1007/s00220-008-0707-y}.

\bibitem{langmann_construction_2022}
E.~Langmann, M.~Noumi and J.~Shiraishi,
\newblock \emph{Construction of eigenfunctions for the elliptic {Ruijsenaars}
  difference operators},
\newblock Commun. Math. Phys. \textbf{391}(3), 901 (2022),
\newblock \doi{10.1007/s00220-021-04195-8}.

\bibitem{konno2023elliptic}
H.~Konno and K.~Oshima,
\newblock \emph{Elliptic quantum toroidal algebra
  ${U}_{q,t,p}(\mathfrak{gl}_{1,tor})$ and affine quiver gauge theories},
\newblock Lett. Math. Phys. \textbf{113}(2), 32 (2023),
\newblock \doi{10.1007/s11005-023-01650-6},
\newblock \eprint{2112.09885}.

\bibitem{BLST24}
A.~Ben~Moussa, J.~Lamers, D.~Serban and A.~Toufik,
\newblock \emph{A solvable non-unitary fermionic long-range model with extended
  symmetry} (2024), \eprint{2404.10164}.

\bibitem{Serban:2004jf}
D.~Serban and M.~Staudacher,
\newblock \emph{Planar {$N = 4$} gauge theory and the {I}nozemtsev long range
  spin chain},
\newblock JHEP \textbf{2004}(06), 001 (2004),
\newblock \doi{10.1088/1126-6708/2004/06/001}.

\bibitem{rej2012review}
A.~Rej,
\newblock \emph{Review of {AdS/CFT} integrability, chapter {I}.3: {L}ong-range
  spin chains},
\newblock Lett. Math. Phys. \textbf{99}, 85 (2012),
\newblock \doi{10.1007/s11005-011-0509-6},
\newblock \eprint{1012.3985}.

\bibitem{de2022lifting}
M.~de~Leeuw and A.~Retore,
\newblock \emph{Lifting integrable models and long-range interactions},
\newblock SciPost Phys. \textbf{15}, 241 (2023),
\newblock \eprint{2206.08390}.

\bibitem{gombor2022wrapping}
T.~Gombor,
\newblock \emph{Wrapping corrections for long-range spin chains},
\newblock Phys. Rev. Lett. \textbf{129}(27), 270201 (2022),
\newblock \doi{10.1103/PhysRevLett.129.270201},
\newblock \eprint{2206.08679}.

\bibitem{koroteev2020quantum}
P.~Koroteev and S.~Shakirov,
\newblock \emph{The quantum {DELL} system},
\newblock Lett. Math. Phys. \textbf{110}(5), 969 (2020),
\newblock \doi{10.1007/s11005-019-01247-y},
\newblock \eprint{1906.10354}.

\bibitem{gorsky2022double}
A.~Gorsky, P.~Koroteev, O.~Koroteeva and S.~Shakirov,
\newblock \emph{Double {I}nozemtsev limits of the quantum {DELL} system},
\newblock Phys. Lett. B \textbf{826}, 136919 (2022),
\newblock \doi{10.1016/j.physletb.2022.136919},
\newblock \eprint{2110.02157}.

\bibitem{lamers2016elliptic}
J.~Lamers,
\newblock \emph{On elliptic quantum integrability: vertex models,
  solid-on-solid models and spin chains},
\newblock Ph.D. thesis, Utrecht University,
\newblock \url{https://dspace.library.uu.nl/handle/1874/333998} (2016).

\bibitem{DLMF}
\emph{{NIST Digital Library of Mathematical Functions}},
\newblock \url{http://dlmf.nist.gov/}, release 1.2.0 of 2024-03-27.

\bibitem{abramowitz1948handbook}
M.~Abramowitz and I.~Stegun,
\newblock \emph{Handbook of mathematical functions with formulas, graphs, and
  mathematical tables}, vol.~55,
\newblock US Government printing office (1948).

\bibitem{whittaker1904course}
E.~Whittaker and G.~Watson,
\newblock \emph{A Course of Modern Analysis: An Introduction to the General
  Theory of Infinite Series and of Analytic Functions, with an Account of the
  Principal Transcendental Functions},
\newblock Cambridge University Press (1902).

\bibitem{TF79}
L.~A. Takhtadzhan and L.~D. Faddeev,
\newblock \emph{The quantum method of the inverse problem and the {H}eisenberg
  {XYZ} model},
\newblock Russ.\ Math.\ Surv. \textbf{34}, 11 (1979),
\newblock \doi{10.1070/RM1979v034n05ABEH003909}.

\bibitem{FELDER1996485}
G.~Felder and A.~Varchenko,
\newblock \emph{Algebraic {B}ethe ansatz for the elliptic quantum group
  {$E_{\tau,\eta}(\mathfrak{sl}_2)$}},
\newblock Nucl. Phys. B \textbf{480}(1), 485 (1996),
\newblock \doi{10.1016/S0550-3213(96)00461-0},
\newblock \eprint{q-alg/9605024}.

\end{thebibliography}

\end{document}